\documentclass[fleqn,usenatbib]{mnras}

\usepackage{newtxtext,newtxmath}

\usepackage[T1]{fontenc}
\usepackage{ae,aecompl}

\usepackage{graphicx}	
\usepackage{amsmath}	
\usepackage{amssymb}	
\usepackage{bm}
\usepackage{adjustbox}

\usepackage{hyperref}

\def\apjl{ApjL}

\newcommand{\ex}{\hat{\bm{x}}}

\newcommand{\ez}{\hat{\bm{z}}}
\newcommand{\bh}{\bm{\hat{b}}}

\newcommand{\pc}{p_c}
\newcommand{\pg}{p_g}

\newcommand{\diff}{\bm{\nabla \cdot }\Big( \kappa \bh \bh \bm{\cdot \nabla} \pc \Big)}
\newcommand{\crheat}{\bm{v_A \cdot \nabla} \pc}
\newcommand{\stream}{\bm{\nabla \cdot} (\bm{v} + \bm{v_A})}

\newcommand{\LT}{\Lambda_T}
\newcommand{\LTC}{\Lambda_{T,c}}

\title[Cosmic-Ray Thermal Instability]{Thermal Instability of Halo Gas Heated by Streaming Cosmic Rays}

\author[Kempski \& Quataert]{
Philipp Kempski,$^{1}$\thanks{E-mail: philipp.kempski@berkeley.edu}
Eliot Quataert$^{1}$
\\
$^{1}$Department of Astronomy and Theoretical Astrophysics Center, University of California, Berkeley, Berkeley, CA 94720, USA}
\date{Accepted XXX. Received YYY; in original form ZZZ}

\pubyear{2019}

\begin{document}
\label{firstpage}
\pagerange{\pageref{firstpage}--\pageref{lastpage}}
\maketitle

\begin{abstract}
 Heating of virialized gas by streaming cosmic rays (CRs) may be energetically important in galaxy halos, groups and clusters. We present a linear thermal stability analysis of plasmas heated by streaming CRs. We separately treat equilibria with and without background gradients, and with and without gravity. We include both CR streaming and diffusion along the magnetic-field direction. Thermal stability depends strongly on the ratio of CR pressure to gas pressure, which determines whether modes are isobaric or isochoric. Modes with $\bm{k \cdot B }\neq 0$ are strongly affected by CR diffusion. When the streaming time is shorter than the CR diffusion time, thermally unstable modes (with $\bm{k \cdot B }\neq 0$) are waves propagating at a speed $\propto$ the Alfv\'en speed.  Halo gas in photoionization equilibrium is thermally stable independent of CR pressure, while gas in collisional ionization equilbrium is unstable for physically realistic parameters. In gravitationally stratified plasmas, the oscillation frequency of thermally overstable modes can be higher in the presence of CR streaming than the buoyancy/free-fall frequency. This may modify the critical $t_{\rm cool}/t_{\rm ff}$ at which multiphase gas is present.  The criterion for convective instability of a stratified, CR-heated medium can be written in the familiar Schwarzschild form $d s_{\rm eff} / d z < 0$, where $s_{\rm eff}$ is an effective entropy involving the gas and CR pressures. We discuss the implications of our results for the thermal evolution and multiphase structure of galaxy halos, groups and clusters.       
\end{abstract}




\section{Introduction}
The short radiative cooling times of virialized gas in galax- ies and clusters suggest that these systems should contain significantly more cool gas at their centers than is observed (\citealt{pf06}). This implies that the hot gas surrounding galaxy halos is also heated, which is thought to come from feedback by star formation and central active galactic nuclei (AGN; e.g., \citealt{gpr2008}). An appreciable fraction of the energy released by AGNs and supernova explosions comes in the form of relativistic cosmic-ray particles (\citealt{mn07}; \citealt{ackermann2013}), which may be important for the dynamics and gas heating in galaxies, halos and clusters (e.g., \citealt{bmv91}; \citealt{loew91}; \citealt{everett08}; \citealt{socrates08}; \citealt{guo08}; \citealt{zweibel_micro}; \citealt{ruszkowski17}; \citealt{zweibel17}; \citealt{ehlert18}).

Cosmic rays are confined in galaxies for times much longer than would be expected from their propagation speed ($\approx$ speed of light), due to scattering off small-scale electromagnetic fluctuations. These fluctuations can be either due to external turbulence, or Alfv\'en waves generated by the cosmic rays themselves. In the self-excitation scenario, wave growth is driven by the cosmic ray streaming instability (\citealt{kp69}): as cosmic rays collectively drift down their pressure gradient, the free energy associated with their velocity anisotropy can excite Alfv\'en waves. Pitch-angle scattering isotropizes the cosmic rays in the frame comoving with the waves, which, in the absence of strong wave damping, limits the CR drift speed to the local Alfv\'en speed (by contrast, the cosmic-ray drift speed can be significantly larger than the Alfv\'en speed in the strong-damping limit; \citealt{skilling71}; \citealt{wiener2013}). In a steady state, the streaming-induced wave growth is balanced by wave damping, so that the energy of the cosmic rays is essentially being transferred to the thermal plasma. This couples the background plasma to the cosmic rays, which heat the gas at a rate $- \bm{v_A \cdot \nabla} p_c$, where $v_A$ is the local Alfv\'en speed and $p_c$ is the CR pressure (\citealt{wentzel1971}).

\cite{guo08}, \cite{jp_1} and \cite{jp_2} showed that this cosmic-ray heating can suppress the cooling catastrophe in clusters for CR pressures that are consistent with observational bounds. Indeed, they found that the required CR pressure (gradient) is small compared to the gas pressure (gradient), as is also found observationally (e.g., \citealt{huber2013}). Whether the same is true in galaxy halos is still unclear (e.g., \citealt{hop19}).



While heating suppresses cooling globally (i.e. on sufficiently long time and length scales) and maintains the hot virialized gas in massive halos in approximate hydrostatic and thermal balance,\footnote{While the hot virialized gas in clusters has a sufficiently high temperature to be seen directly in emission, the emission from virialized gas in the CGM is too faint for current telescopes. Nevertheless, hot virialized gas is expected to be present in halos of mass $\gtrsim 10^{11.5} M_{\odot}$ (\citealt{bd03}; \citealt{dekel09}).} there is strong observational evidence for cold gas in the halos of galaxies. Cool gas is present both in the circumgalactic medium (CGM) of massive and Milky-Way-like galaxies, and in the intracluster medium (ICM). In the ICM, detailed spatially-resolved observations (that use both atomic and molecular transitions, e.g., \citealt{salome2006}, \citealt{cavagnolo2009}) indicate the presence of cold-gas filaments embedded within the otherwise hot, virialized gas, which constitutes most of the intracluster gas mass. In the CGM, the cool-gas morphology is less certain (i.e., it could be filamentary or volume-filling), and the cold gas mass may comprise a significant fraction of the total halo gas mass. Indeed, observations of the CGM using Ly $\alpha$ emission and quasar absorption lines (\citealt{werk2013}; \citealt{stocke2013}; \citealt{cantalupo2014}; \citealt{hennawi2015}; \citealt{bowen2016}; \citealt{cai2017}) suggest the presence of multiphase gas along most lines of sight  (suggesting that the cold phase may permeate the CGM, instead of forming a filamentary structure).  

The origin of the cold gas remains uncertain. It could be gas elevated into the halo by galactic winds. However, how the cold gas in high-velocity galactic winds is produced and entrained remains uncertain (\citealt{sb2015}; \citealt{thompson2016}; \citealt{zhang2017}). Alternatively (or, in addition), the cold phase may be produced in situ via thermal instability. Thermal instability is commonly linked to the existence of multiphase gas in the interstellar medium (\citealt{field65}) and has been studied in the context of galaxy halos and clusters using a number of simulations and models (\citealt{nulsen1986}; \citealt{binney2009}; \citealt{msqp12}; \citealt{smqp12}; \citealt{voit2015}; \citealt{meece2015}; \citealt{voit2017}; \citealt{voit2018}). These simulations suggest that the condensation of cold gas via thermal instability can occur if the ratio of the cooling time to the free-fall time is sufficiently small. Typically they find that $t_{\rm cool} / t_{\rm ff} \lesssim 10$, however, this value may depend on the size of the initial perturbations (\citealt{pizzolato2005}; \citealt{singh2015}; \citealt{csq19}) and whether magnetic fields are included (\citealt{ji18}). The connection between $t_{\rm cool} / t_{\rm ff} \lesssim 10$ and the existence of multiphase structure has been partly born out by the cluster observations of \cite{mcdonald2010}, but in a more recent sample of 56 clusters observed by the \textit{Chandra X-ray Observatory}, cold gas is present even when $t_{\rm cool} / t_{\rm ff} \gtrsim 10$  (\citealt{hogan17}).  

The purpose of this paper is to understand the thermal stability of systems heated by streaming cosmic rays, which may be an important heating mechanism in galaxy halos. We first present order-of-magnitude estimates showing that heating due to streaming CRs may be important for a wide range of halo masses. We then perform a linear stability analysis, in which we take into account both CR streaming and diffusion, and we look at equilibria with and without gravity. While we find that explicitly including gravity is not very important for thermal instability growth rates, it can transform thermal instability into a convective instability driven by buoyancy.\footnote{This is a rather unsurprising side result of our analysis, because thermal and convective stability are closely linked (\citealt{balbus95}).}

The thermal stability of systems with heating by streaming CRs was first considered heuristically in the context of a cooling flow by \cite{loew91}. \cite{pfrommer13} and \cite{wiener2013b} then studied thermal instability with CR heating by assuming that the CR pressure ($p_c$) and gas density ($\rho$) follow the adiabatic relation $p_c \propto \rho^{\gamma_{c}}$. In this work, we instead explicitly include the evolution equation for the CR pressure, which is in general not consistent with adiabaticity. Cosmic rays are adiabatic only for modes propagating perpendicular to the magnetic field (see Section \ref{sec:uni_disp}), but even then we show that correctly perturbing the CR heating produces results that are different from the heuristic calculation in \cite{pfrommer13}. We also extend previous work by studying the impact of CR diffusion on thermal instability, and we study the instability in different background equilibria.

The remainder of this paper is organised as follows. We introduce the gas--CR equations in Section \ref{sec:eqns}. In Section \ref{sec:balcool} we argue that cosmic-ray heating may be important in galactic halos. The linear thermal stability of CR heating is derived in Sections \ref{sec:uniform}, \ref{sec:nonuni} and \ref{sec:gravity}. We solve the perturbed linearised equations in a uniform medium without gravity in Section \ref{sec:uniform}. We introduce gas and CR background gradients in Section \ref{sec:nonuni} and consider gravitationally stratified equilibria in Section \ref{sec:gravity}. In the latter case, we also obtain a criterion for convective instability. We summarize our results and discuss their implications for the multiphase structure of galaxy halos in Section \ref{sec:conclusions}. 

We derive estimates for the (global) ratio of CR to thermal pressure in galaxy halos in Appendix \ref{app:eta}. A heuristic description of the impact of CR diffusion on thermal instability is provided in Appendix \ref{app_diff}. We show the linearised perturbed equations of a CR-heated background in Appendix \ref{app:nonuni}. In Appendix \ref{app:1D_val}, we explain why a 1-dimensional calculation (see Section \ref{sec:nonuni_1d}) gives the correct eigenfrequency of the gas entropy mode in a CR-heated background. Finally, we derive an approximate growth rate for the convective instability in a gravitationally stratified medium in Appendix \ref{app:conv}.

\section{Equations and Timescales }\label{sec:eqns}
\subsection{Gas--CR Equations}
We consider a thermal plasma interacting with a population of relativistic cosmic rays. We model the system by including CR heating and the CR pressure force in the equations of ideal MHD. This results in the following coupled differential equations,
\begin{equation} \label{eq:cont}
\frac{\partial \rho}{\partial t} + \nabla \cdot ( \rho \bm{v} ) = 0,
\end{equation}
\begin{equation}\label{eq:mom}
\rho \frac{d \bm{v}}{d t} = - \bm{\nabla} \Big( p_g + p_c + \frac{B^2}{ 8 \pi} \Big) + \frac{\bm{B \cdot \nabla B}}{4 \pi} + \rho \bm{g},
\end{equation}
\begin{equation}\label{eq:ind}
\frac{\partial \bm{B}}{\partial t} = \bm{\nabla \times} (\bm{v \times B}),
\end{equation}
\begin{equation}\label{eq:s}
\rho T \frac{ds}{dt} = \mathcal{H} -\crheat - \rho ^2 \Lambda (T),
\end{equation}
\begin{equation} \label{eq:pc}
\frac{dp_c}{dt} = -\frac{4}{3}p_c \stream - \crheat + \diff
\end{equation}
where  $\bm{v}$ is the gas velocity, $\rho$ is the gas density, $\pg$ and $\pc$ are the gas and CR pressures respectively, $\bm{B}$ is the magnetic field (with unit vector along $\bh$), $\bm{g}$ is the acceleration due to gravity, $s=k_{\rm B} \ln(p/\rho^{\gamma}) / (\gamma -1)m_{\rm H}$ is the gas entropy per unit mass, $\Lambda(T)$ is the temperature-dependent cooling function, and $\mathcal{H}$ is an unspecified heating rate (which we set to 0 except in Section \ref{sec:uniform}). $d / dt \equiv \partial / \partial t + \bm{v \cdot \nabla}$ denotes a total (Lagrangian) time derivative. We assume that cosmic rays stream down their pressure gradient at the Alfv\'en velocity $\bm{v_A} = \bm{B} / \sqrt{4\pi\rho}$, and we also include CR diffusion along the magnetic field, for which we assume a constant diffusion coefficient $\kappa$. We note that formally CRs stream with velocity $\bm{v_{\rm st}} = -{\rm sgn}(\bm{\hat{b} \cdot \nabla}p_c) \bm{v_A}$. This ensures that cosmic rays stream along the magnetic field down their pressure gradient and makes the CR heating term $-\bm{v_{\rm st} \cdot \nabla}p_c$ positive definite. In our linear stability analysis cosmic rays stream at $\bm{v_A}$, as we consider background equilibria which satisfy $-\bm{v_A \cdot \nabla} p_c >0$ (see footnote \ref{ftn:pos_def} in Section \ref{sec:uni_eq} for how this is achieved in a uniform background).  

\subsection{CR Transport: Streaming versus Diffusion} \label{sec:stream_diff}
The interplay between cosmic-ray streaming and diffusion calls for some further discussion. In the self-confinement picture, the importance of streaming versus diffusion is intimately tied to the saturation of the streaming instability (\citealt{kc71}; \citealt{skilling71}; \citealt{wiener2013}). In the limit of weak damping, the excited Alfv\'en waves can grow to large amplitudes (compared to when significant damping is present, see next paragraph)  until the resultant rapid pitch-angle scattering isotropizes the CRs in the frame of the waves. In this scenario, the cosmic rays are advected down their pressure gradient at the Alfv\'en speed relative to the thermal plasma, with no diffusive contribution (we neglect diffusion due to external turbulence). This is tantamount to setting $\kappa=0$ in the above equations. Note that the term $- \bm{v_A \cdot \nabla} p_c$ in equation \ref{eq:s} is then positive definite, because in the limit of self-excited Alfv\'en waves only (no background turbulence) energy flows from the CRs to the gas (mediated by Alfv\'en waves), but not vice-versa.

In the opposite limit of significant wave damping, the waves generated by the streaming instability saturate at lower amplitudes. As a result, the CR pitch-angle scattering rate is reduced, and the cosmic-ray momenta do not become fully isotropic in the Alfv\'en-comoving frame. In this case, the cosmic-ray bulk motion deviates from pure streaming at $\bm{v_A}$ and $\kappa$ will generally be nonzero. The diffusion coefficient will depend on how the waves are damped. Quite notably, for many of the known damping mechanisms (e.g. turbulent, ambipolar and linear Landau damping), the diffusion term ends up not being diffusive at all (\citealt{skilling71}; \citealt{wiener2013}; \citealt{wiener18}). Instead, it has the form of an advective flux (streaming) and the cosmic rays essentially stream down their pressure gradient at super-Alfv\'enic speeds. This, however, is not always true (e.g., when non-linear Landau damping is dominant and/or if there are external sources of cosmic-ray scattering distinct from self-excited Alfv\'en waves). For this reason, we keep the diffusion term in our equations (with constant $\kappa$ for simplicity). We do not consider super-Alfv\'enic streaming in this work, as the dependence of super-Alfv\'enic streaming velocities on other fluid quantities is uncertain.

\subsection{Dimensionless Parameters and Characteristic Frequencies}
We define the ratio of CR pressure to gas pressure,
\begin{equation} \label{eq:eta}
 \eta \equiv \frac{p_c}{p_g}   ,
\end{equation}
and the ratio of thermal to magnetic pressure, 
\begin{equation} \label{eq:beta}
    \beta \equiv \frac{8\pi p_g} {B^2} .
\end{equation}
We also write the logarithmic slope of the cooling function as
\begin{equation}
    \LT \equiv  \frac{\partial \ln \Lambda}{\partial \ln T}.
\end{equation}
There are a number of timescales that characterise the problem. We define the cooling frequency,
\begin{equation} \label{eq:wc}
\omega_c \equiv \frac{\rho ^2 \Lambda}{p_g}; 
\end{equation}
the wavenumber ($k$) dependent sound frequency (with $c_s$ being the adiabatic sound speed),
\begin{equation} \label{eq:ws}
    \omega_s \equiv k c_s ;
\end{equation}
 the Alfv\'en \textit{and} CR-heating frequency,
\begin{equation}\label{eq:wa}
\omega_a \equiv  \bm{k \cdot v_A} ;
\end{equation}
the cosmic-ray diffusion frequency,
\begin{equation}\label{eq:wd}
\omega_d \equiv \kappa \  (\bm{\hat{b} \cdot k})^2 ;
\end{equation}
and the free-fall frequency,
\begin{equation}\label{eq:wff}
    \omega_{\rm ff} \equiv \frac{g}{c_s}.
\end{equation}
We stress that $\omega_a$ characterizes both the perturbed magnetic tension (its usual meaning) and the perturbed CR heating $- \bm{v_A \cdot \nabla}p_c$. Throughout our linear stability calculation in Sections \ref{sec:uniform}, \ref{sec:nonuni} and \ref{sec:gravity}, we focus on local perturbations ($kH \gg 1$, $H$ being a characteristic background length scale), which for our application considered in Section \ref{sec:balcool} corresponds to
\begin{equation} \label{eq:w_s_vs_cff}
    \omega_s \gg \omega_c, \omega_{\rm ff},
\end{equation}
and
\begin{equation} \label{eq:w_a_vs_cff}
    \omega_a \gg \omega_c
\end{equation}
(unless $\mathbf{k \cdot B}=0$, in which case eq. \ref{eq:w_a_vs_cff} need not be satisfied). In the CR-heated background, $kH \gtrsim 1$ corresponds to $\omega_a \gtrsim \omega_c \eta^{-1}$ (Section \ref{sec:nonuni_eq}). We find that thermal instability growth rates do not depend significantly on wavenumber $k$, provided that $\omega_a > \omega_c$ ($\omega_a \gtrsim \omega_c \eta^{-1}$) in the uniform (CR-heated) background. 

As our fiducial set of parameters, we choose $\omega_a = 10^3 \omega_c$ (which corresponds to fairly high $k$, but such high $k$ is necessary for the CR-heated background if we want to consider $\eta > 0.01$), $\beta = 100$ and, when we include gravity in Section \ref{sec:gravity}, $\omega_{\rm ff} = 20 \omega_c$. $\omega_{\rm ff} \gtrsim 10 \omega_c$ is motivated by observations of hot gas in groups and clusters (\citealt{mcdonald2010}; \citealt{hogan17}), which largely satisfy this constraint. We stress that this choice of $\omega_{\rm ff} \gtrsim 10 \omega_c$ is motivated by  halo gas specifically, but need not be true in other applications. We show how smaller $\beta$ and $\omega_a$ affect our results in Figure \ref{fig:uni}.

\section{Heating by Cosmic Rays in Galaxy Halos} \label{sec:balcool} 
 \begin{figure} 
  \centering
    \begin{minipage}[b]{0.45\textwidth}
    \includegraphics[width=\textwidth]{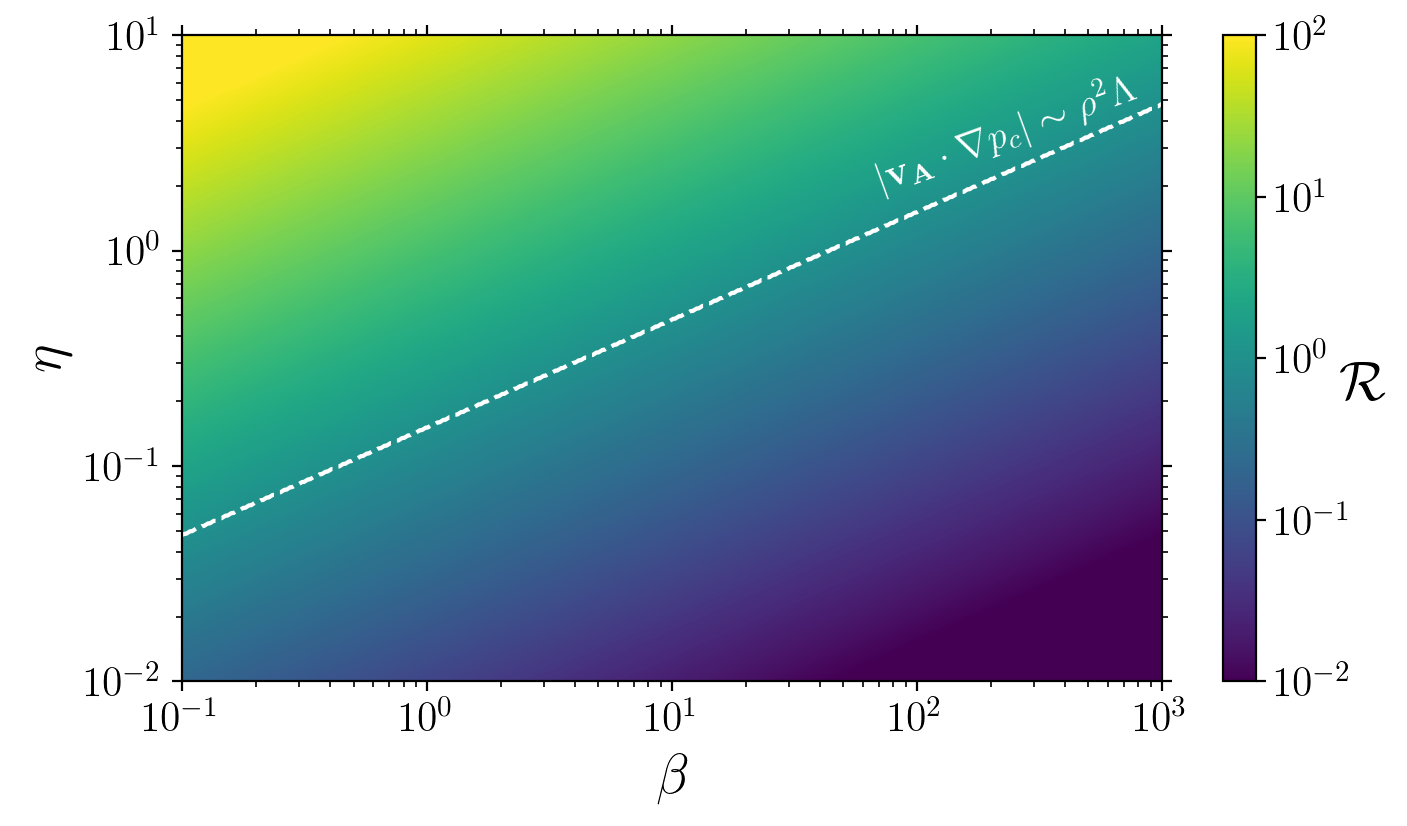}
  \end{minipage}
  \begin{minipage}[b]{0.45\textwidth}
    \includegraphics[width=\textwidth]{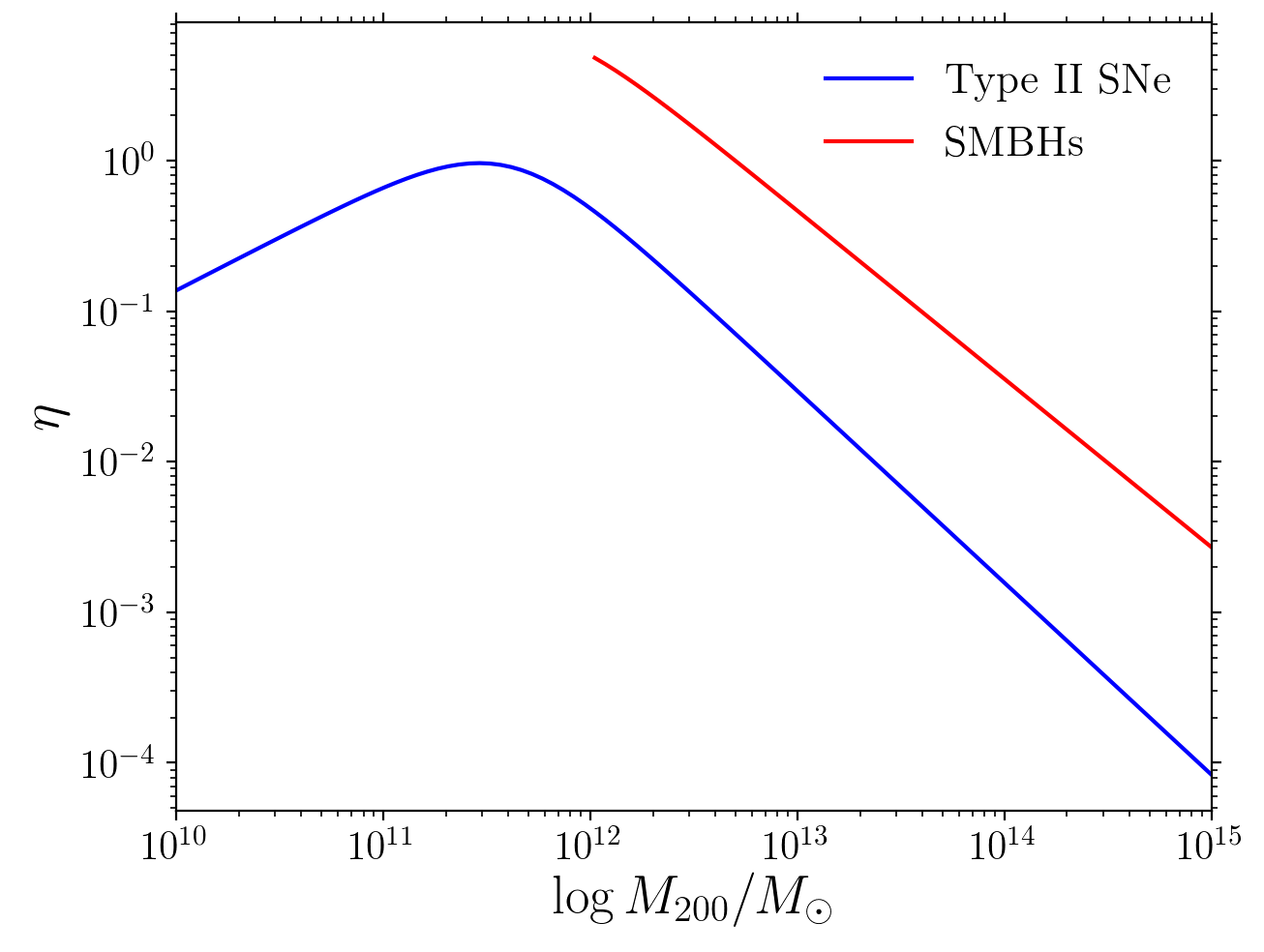}
  \end{minipage}
  \caption{ \textbf{Top:} CR heating versus cooling as a function of $\eta$ (eq. \ref{eq:eta}) and $\beta$ (eq. \ref{eq:beta}). $\mathcal{R}$ is the ratio of CR heating to radiative cooling (eq. \ref{eq:R}; here we use $\omega_{\rm ff} = 20 \omega_c$ and $H_c = 3 H$), and increases with increasing CR pressure fraction $\eta$ and with decreasing $\beta$. The white dashed line indicates the approximate region where  cosmic-ray heating is comparable to cooling ($\mathcal{R} \sim 1$). \textbf{Bottom:} Order-of-magnitude estimate of the CR pressure fraction as a function of halo mass (see Appendix \ref{app:eta}). We separately consider cosmic rays injected into the halo by Type II SNe and central SMBHs.  We include this plot to motivate that significant cosmic ray pressures are plausible for a wide range of halo masses, especially in $M \approx 10^{12} M_{\odot}$ halos (large CR pressures, $\eta \gtrsim 1$, have also been found in cosmological simulations, see e.g. \citealt{hop19}). Together, the two panels suggest that there may be significant cosmic-ray heating in galaxy halos for a wide range of halo masses.  \label{fig:heat_vs_cool}} 
\end{figure}

Before we look at the thermal stability of CR heating, we check under what conditions thermal balance between CR heating and radiative cooling,
\begin{equation} \label{eq:heat_bal}
    - \crheat = \rho^2 \Lambda(T),
\end{equation}
may occur in galaxy halos. Our estimates presented in this section suggest that heating by cosmic rays can be important for a broad range of values of $\eta$ and $\beta$ (see Figure \ref{fig:heat_vs_cool}).

We quantify the importance of cosmic-ray heating by defining
\begin{equation} \label{eq:Rdef}
    \mathcal{R} \equiv  \frac{ |\bm{v_A \cdot \nabla} p_c|  }{\rho^2 \Lambda} \sim \frac{ v_A \pc /H_{c}}  { \omega_c \pc \eta^{-1}},
\end{equation}
where we used our definition of the cooling frequency \eqref{eq:wc}. $H_c$ is the CR pressure scale height. Using definition \ref{eq:wff} and $\beta \approx (c_s / v_A)^2$, we obtain
\begin{equation} \label{eq:R}
    \mathcal{R} \sim \frac{\eta}{\sqrt{\beta}}  \frac{H}{H_c} \frac{ \omega_{\rm ff}}{\omega_c},
\end{equation}
where $H = c_s^2 /g$. When gas pressure dominates, $H$ is approximately equal to the gas pressure scale height $H_g$.

Our estimate for $\mathcal{R}$ as a function of $\eta$ and $\beta$ is plotted in the top panel of Figure \ref{fig:heat_vs_cool} for $\omega_{\rm ff}/\omega_c = 20$. We choose a fairly extended CR profile, with $H_{c} / H = 3$. The dashed white line indicates the approximate region where heating by cosmic rays is comparable to cooling, i.e. $\mathcal{R} \sim 1$.

The top panel of Figure \ref{fig:heat_vs_cool} suggests that heating by cosmic rays may be important for a wide range of $\eta$ and $\beta$. There is some evidence, from both observations and theory/simulations, suggesting that galaxy halos may often reside above/around the white dashed line (where CR heating is important). While significant CR pressures were measured in the Milky Way (\citealt{boulares1990}) and nearby starburst galaxies (\citealt{paglione2012}), observations of cosmic rays and magnetic fields in galaxy halos are challenging and sparse. Nevertheless, there are some constraints that come from synchrotron emission and Faraday rotation measurements along quasar sightlines. Synchrotron emission measurements suggest that cosmic rays and magnetic fields have significantly larger scale heights than the thermal gas (\citealt{Beck2015}). There is also evidence for strong magnetic fields ($1-10 {\rm \mu G}$) that may extend far out (tens of kpc) into the halo (\citealt{morakrause2013}; \citealt{bernet2013}). As a result, it is plausible that there are regions in the halo where $\eta$ is large (e.g., $\sim 1$) and/or $\beta$ is relatively small (e.g., $\lesssim 10$). Under such conditions, equation \ref{eq:R} and the top panel of Figure \ref{fig:heat_vs_cool} suggest that there may be significant CR heating. 

Recent cosmological zoom-in simulations with cosmic rays strengthen the claim that CR pressure can be important (even dominant) in galaxy halos (\citealt{hop19}). This is broadly consistent with our estimate for the CR pressure fraction $\eta$, which we show as a function of halo mass in the bottom panel of Figure \ref{fig:heat_vs_cool} (the calculation can be found in Appendix \ref{app:eta}). We separately consider the injection of cosmic rays by Type II Supernovae and central Supermassive Black Holes (SMBHs), and we estimate the total energy of cosmic rays out to the virial radius. We assume the (broken power law) stellar mass -- halo mass relation from \cite{moster2013} and the SMBH mass -- total stellar mass relation (for ellipticals) from \cite{rv2015}.  Comparing the CR energy to the total thermal energy within the virial radius yields the lower panel of Figure \ref{fig:heat_vs_cool}. We find that CR pressure should be significant for a broad range of halo masses and most important in halos of mass $\approx 10^{12} M_{\odot}$, consistent with \cite{hop19}. 


\section{Cosmic-Ray Thermal Instability in a Uniform Medium} \label{sec:uniform}
Before we analyse equilibria in which CR heating balances cooling (due to a finite background CR pressure gradient), we look at the simpler case of a uniform background. This setup is particularly relevant for cases where CR heating is not the dominant heating process, but can nevertheless affect the evolution of entropy perturbations (photoionization equilibrium is one such example). The uniform-medium calculation does not capture (slight) modifications to the thermal instability that come from a background CR pressure gradient, but in many ways it produces results that are very similar to the non-uniform medium calculation. For example, the thermal instability growth rates have an almost identical dependence on $\eta$ and CR diffusion. As a result, many of the conclusions drawn here will still be valid in the calculation with background CR heating.

We perform a linear stability calculation of the equations described in Section \ref{sec:eqns}. All perturbed quantities are assumed to vary as $\delta Q(\bm{r}, t) \propto \exp \Big[i \bm{k \cdot r} - i \omega t \Big]$. Throughout this (and the next) section, we also ignore gravity, i.e. we set $\bm{g} = 0$ (we include gravity in Section \ref{sec:gravity}).

\subsection{Equilibrium} \label{sec:uni_eq}
We consider an equilibrium with
\begin{equation}
    \mathcal{H} = \rho^2 \Lambda(T),
\end{equation}
where $\mathcal{H}$ is an unspecified heating rate, which is set to balance cooling (i.e. $\mathcal{H} \gg -\bm{v_A \cdot \nabla} p_c$). Equilibrium CR heating is considered negligible, and the CR heating term only enters in the perturbed equations. All background fluid variables are assumed to be spatially constant. Without loss of generality, we consider a vertical magnetic field, $\bm{B} = B \ez$. This equilibrium has the advantage that there are no background gradients in our linear stability analysis.\footnote{For $\nabla p_c$ to have a well-defined sign in our linear stability analysis, so that $- \bm{v_A \cdot \nabla} p_c$ is positive definite, $p_{c}$ cannot be completely uniform. We therefore need a small background CR pressure gradient and to this end, we write $- \bm{v_A \cdot \nabla} p_{c} =  \epsilon \rho^2 \Lambda $. In our linear calculation we can then still drop any background gradients if we adopt the ordering $1 \gg \epsilon \gg \delta Q /Q$ for any quantity $Q$. Under this ordering, we can essentially treat the equlibrium $\rho$, $\pg$ and $\pc$ as uniform. We note, however, that this approach breaks down when $\delta p_c / p_c > (k H_{c})^{-1}$, as the perturbations are large enough to flatten out the CR pressure distribution and decouple the cosmic rays from the gas. In the small-background-gradient limit that is the assumption in our uniform medium calculation, this can in practice happen at small $\delta p_c / p_c.$  \label{ftn:pos_def}} Moreover, treating $\kappa$ as a constant (and not a function of $B, \ p_c$ and other fluid variables) is exact to linear order in a uniform background.
\subsection{Linearised Equations}
We ignore perturbations of $\mathcal{H}$, i.e. we set $\delta \mathcal{H}=0$ (generalization to finite $\delta \mathcal{H}$ is straightforward). We do, however, perturb the cosmic-ray heating term. The linearised perturbed versions of equations \ref{eq:cont}--\ref{eq:pc} are
\begin{equation} \label{eq:uni_drho}
    \frac{\delta \rho}{\rho} = -i \bm{k \cdot \xi},
\end{equation}
\begin{equation}  \label{eq:uni_dxi}
    -\rho \omega^2 \bm{\xi} = -i \bm{k} \Big( \delta p_c + \delta p_g + \frac{\bm{B \cdot \delta B}}{4\pi} \Big) + i \frac{\bm{(B\cdot k) \delta B}}{4\pi},
\end{equation}
\begin{equation} \label{eq:uni_dB}
    \bm{\delta B} = i(\bm{B \cdot k})\bm{\xi} - i \bm{B}(\bm{k \cdot \xi}),
\end{equation}

\begin{equation} \label{eq:uni_pg}
    \frac{\delta p_g}{p_g} \Big( \frac{\omega}{\gamma - 1} + i \omega_c \LT   \Big) - \omega_a \frac{\delta p_c}{p_g} = \frac{\delta \rho}{\rho} \Big( \omega \frac{\gamma}{\gamma-1} - i \omega_c  (2 - \LT )  \Big),
\end{equation}

\begin{equation} \label{eq:uni_pc}
    \frac{\delta p_c}{p_g} (\omega - \omega_a + i \omega_d) = \frac{\delta \rho}{\rho} \eta \Big( \frac{4}{3}\omega - \frac{2}{3} \omega_a \Big).
\end{equation}
\subsection{Dispersion Relation}
\label{sec:uni_disp}

We find the exact solutions to \eqref{eq:uni_drho}--\eqref{eq:uni_pc} by numerically solving for the matrix eigenvalues using MATLAB (because the complete dispersion relation is long and not very enlightening, we do not write it down explicitly). We filter out Alfv\'en waves, which decouple and do not affect thermal instability, by restricting $\bm{\xi}$, $\bm{\delta B}$ and $\bm{k}$ to lie in the $xz$-plane. The exact gas entropy eigenmode that can be derived from \eqref{eq:uni_drho}--\eqref{eq:uni_pc} is necessary for studying thermal instability at low $\beta$. However, we find that our results depend only mildly on $\beta$ for $\beta \gtrsim 3$ (see middle panels of Figure \ref{fig:uni}). In the high-$\beta$ regime the equations simplify considerably, as the CR and gas pressures satisfy the approximate pressure balance $\delta p_c \approx - \delta p_g$. Equations \ref{eq:uni_pg} and \ref{eq:uni_pc} then decouple from the rest (cf. thermal instability is associated with the entropy mode in standard hydrodynamics/MHD) and we end up with a quadratic dispersion relation:
\begin{gather}
\begin{aligned} \label{eq:dispersion_quad_uni}
     0 = \eta \Big( \frac{4}{3}\omega - \frac{2}{3} \omega_a \Big) \Big( \frac{3}{2}\omega +  \omega_a + i \omega_c \LT \Big) \ + \\ \Big( \omega - \omega_a + i \omega_d \Big) \Big( \frac{5}{2} \omega - i \omega_c (2 - \LT  )   \Big) .
    \end{aligned}
\end{gather}
Note that $\omega_a$ here is due to the perturbed CR heating, and not due to the magnetic tension or pressure forces; the latter are 0 in the approximation used here. We show the calculated growth rate as a function of $\eta$ in the top panels of Figure \ref{fig:uni}, focusing mainly on our fiducial parameters, $\omega_a = 10^{3}\omega_c$ and $\beta=100$. The solution of equation \ref{eq:dispersion_quad_uni} is not explicitly plotted, as it agrees almost perfectly with the exact solution at $\beta=100$ and would not be visible (the second solution to equation \ref{eq:dispersion_quad_uni}, associated with the CR entropy mode, is not shown as it is stable for all $\eta$).  The left plot shows the growth rate for different cooling curve slopes $\LT$. For $\LT=-1$, we also show how our results change for $\omega_a = 10 \omega_c$ (green dashed line; no visible change), $\omega_a = \omega_c$ (green dash-dotted line) and $\omega_a = 0$ (i.e. $\bm{k \cdot B}= 0$; green dotted line). The middle panel shows how the $\beta=100$ growth rate ($\approx \beta \rightarrow \infty$ growth rate; blue line, $\LT =-1$) compares to growth rates at smaller $\beta$. We see that the agreement with, e.g., the $\beta = 3$ calculation is still remarkably good. The right panel shows the effects of diffusion (again for $\LT = -1$) for modes with $\omega_d = 0$ (blue), $\omega_a \gg \omega_d \gg \omega_c$ (orange) and $\omega_d \gg \omega_a$ (green). For more discussion of the effects of diffusion, see Section \ref{sec:uni_diff} and Appendix \ref{app_diff}. 

In the case $\omega_a = 0$ (due to $\bm{k\cdot B} = 0$; green dotted line in left panel) and $\omega_d = 0$ (no diffusion), equation \ref{eq:uni_pc} reduces to an adiabatic relation between $\delta p_c$ and $\delta \rho$, with adiabatic index $4/3$. Our perpendicular-modes calculation is therefore the closest to the calculation in \cite{pfrommer13}, who assumed an adiabatic relation between CR pressure and gas density. However, our results are different, as the heuristic calculation in \cite{pfrommer13} is not accurate: in particular, their perturbed CR heating is not correct.\footnote{For perpendicular modes ($\omega_a =0$), the perturbed CR heating is $\bm{\delta v_A \cdot \nabla} P_c$, which is 0 in a uniform background. In the adiabatic calculation in \cite{pfrommer13}, the CR heating is incorrectly assumed to scale as $H_{\rm CR} \propto \rho^{\gamma_c + 1/3 - 1/2}$ and contributes to thermal instability as long as there are density perturbations. This heating term is dominated by an assumed dependence $H_{\rm CR} \propto \delta p_c \propto \rho^{\gamma_c}$; in fact, because CRs are adiabatic only for perpendicular modes with $\omega_a = 0$, there is no contribution to $H_{\rm CR}$ from $\delta p_c$. Moreover, for modes with $\bm{k \cdot B} \neq 0$, perturbations to $H_{\rm CR} \propto \bm{v_A \cdot \nabla} \delta p_c \propto \omega_a \delta p_c$ primarily contribute to an oscillatory response, not a change to the growth rate. This is also not captured in the heuristic calculation in \cite{pfrommer13}.}

\subsubsection{Effect of CR Streaming on Entropy Modes}
\label{sec:uni_streaming}
Before discussing the thermally unstable modes driven by cooling in more detail, we first consider the effect of CR streaming on the entropy modes, which becomes clear if we ignore cooling and CR diffusion in \eqref{eq:dispersion_quad_uni}, i.e. consider $\omega_c = \omega_d = 0$. The dispersion relation then becomes:
\begin{equation} \label{eq:uni_dr_stram}
     \eta \Big( \frac{4}{3}\omega - \frac{2}{3}\omega_a   \Big) \Big( \frac{3}{2}\omega +  \omega_a \Big) + \frac{5}{2}  \omega \Big(    \omega - \omega_a \Big)  = 0.
\end{equation}
This dispersion relation is in fact a statement of pressure balance and can be obtained by setting $\delta p_c + \delta p_g = 0$ (without cooling and CR diffusion). When CR pressure is negligible ($\eta \rightarrow 0$), we see that the two solutions are the ordinary MHD gas entropy mode, with $\omega = 0$ (as CR heating is negligible), and the CR entropy mode, which due to the perturbed work done by the CRs ($- \bm{v_A \cdot \nabla} \delta p_c$)\footnote{Due to the ``-", $- \bm{v_A \cdot \nabla}p_c$ is actually positive definite, possibly suggesting that the CRs gain energy according to eq. \ref{eq:pc}. However, when the CR energy equation is rewritten in the conservative form, 
$$\frac{\partial p_c}{\partial t} + \frac{4}{3} \bm{\nabla \cdot} \Big( (\bm{v}+ \bm{v_A})p_c \Big)  =  \frac{1}{3} (\bm{v} + \bm{v_A}) \bm{\cdot \nabla} p_c, $$
it becomes clear that this term is in fact associated with the work done by the CRs on the Alfv\'en waves (and hence the gas). Note that the CR energy is $E_c = 3 p_c$.
} has a frequency $\omega = \omega_a$. 

When CR pressure dominates ($\eta \gg 1$), the CR entropy mode frequency is $\omega =  \omega_a /2$, as can also be seen directly from equation \ref{eq:uni_pc} (with $\omega_d = 0$). This comes directly from the CR compressibility term $-(4/3) p_c \bm{\nabla \cdot} (\bm{v}+\bm{\delta v_A})$, which at large CR pressures is more important for the CR entropy mode evolution than the work done by the CRs on the gas (which is related to the term $- \bm{v_A \cdot \nabla} \delta p_c$). The gas entropy mode at large $\eta$ is isochoric ($|\delta p_c / p_g| \approx |\delta p_g / p_g| \gg |\delta \rho / \rho|$; see Section \ref{sec:perts}). CR heating then dominates the evolution of gas-pressure oscillations (LHS of eq. \ref{eq:uni_pg}) and the oscillations occur at a frequency $\omega = - (2/3) \omega_a$. 

Thus, CR streaming always gives rise to an oscillatory frequency $\mathcal{O}(\omega_a)$ in the CR entropy mode, while in the gas entropy mode CR heating introduces oscillations as long as $\eta$ is finite, and the oscillation frequency approaches $\mathcal{O}(\omega_a)$ once $\eta \sim 1$. Note that while in the classic calculations of thermal instability (e.g., \citealt{field65}) the entropy mode is overstable just due to gravity (rather than purely growing when there is no gravity), thermal instability modes are overstable even without gravity when there is a finite CR pressure. In particular, in the presence of CR heating thermally unstable modes are waves propagating at a speed $\propto v_A$.

\subsection{Density versus Temperature Perturbations} \label{sec:perts}
Equation \ref{eq:uni_pc} (and $\delta p_c \approx - \delta p_g$) offers insight into the relative importance of $\delta p_g$ and $\delta \rho$ for driving thermal instability. This turns out to depend primarily on the CR pressure fraction $\eta$, due to the coupling of $\delta p_c$ and $\delta \rho$ via equation \ref{eq:uni_pc}. Typically, we have that:\footnote{The exceptions to this are if $\omega_d \gg \omega_a$ (so that diffusion wipes out the CR pressure perturbation), or $\omega =  \omega_a - i \omega_d$ or $\omega =  \omega_a /2$, which are the CR entropy modes at small and large $\eta$ respectively, see Section \ref{sec:uni_streaming}.}
\begin{equation} \label{eq:perts}
    | \delta p_g / p_g | \approx |\delta p_c / p_g| \sim \eta | \delta \rho / \rho| ,
\end{equation}
so that perturbations are essentially isobaric for $\eta \ll 1$ and isochoric when $\eta \gg 1$ (large CR pressure stiffens the gas). For large $\omega_d$, CR pressure perturbations are suppressed because they are smoothed out by diffusion, and perturbations are isobaric up to larger $\eta$.

\subsection{Asymptotic Limits} \label{sec:uni_asymptotes}
\begin{figure*} 
  \centering
  \begin{minipage}[b]{0.32\textwidth}
    \includegraphics[width=\textwidth]{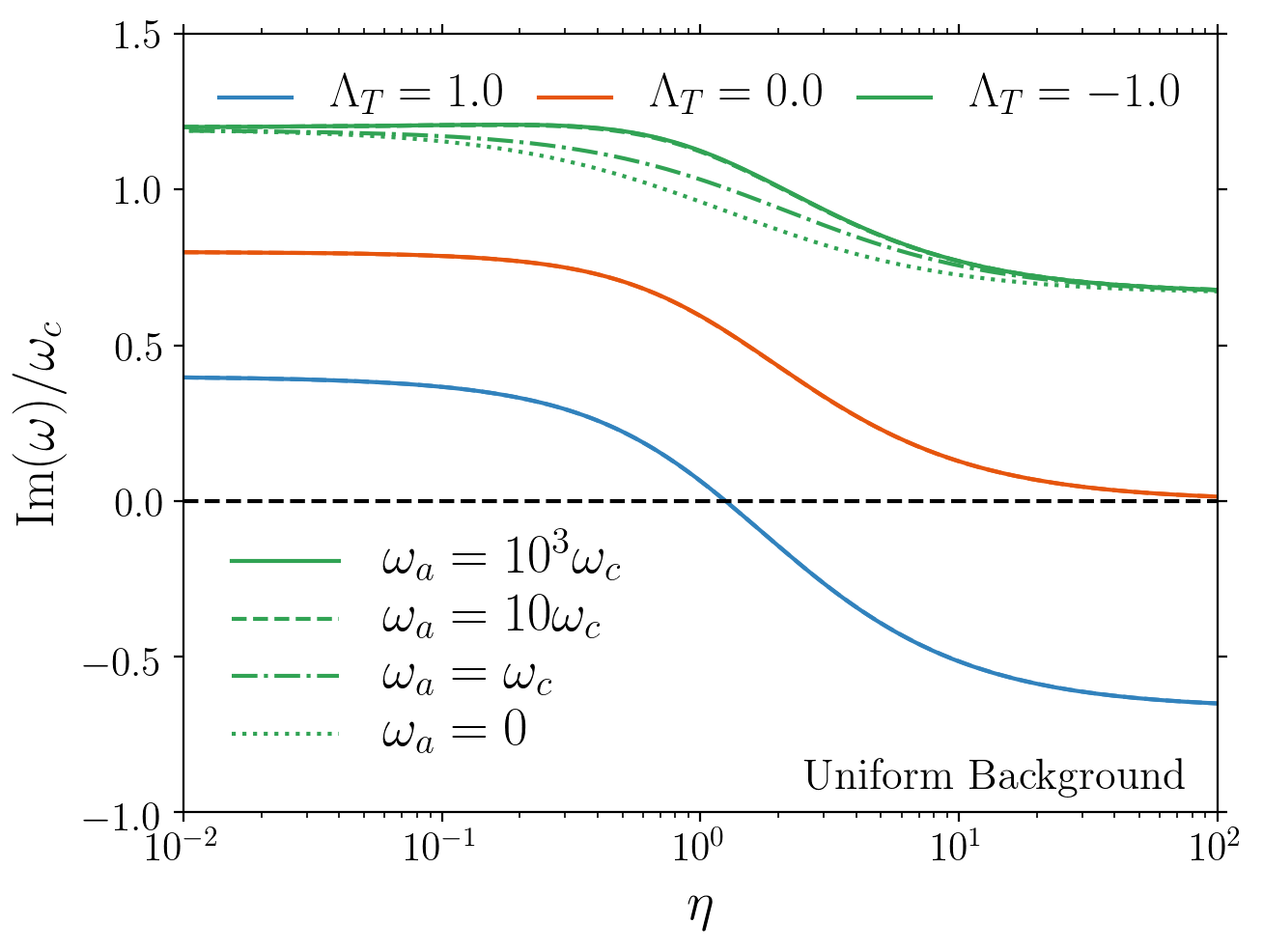}
  \end{minipage}
  \begin{minipage}[b]{0.32\textwidth}
    \includegraphics[width=\textwidth]{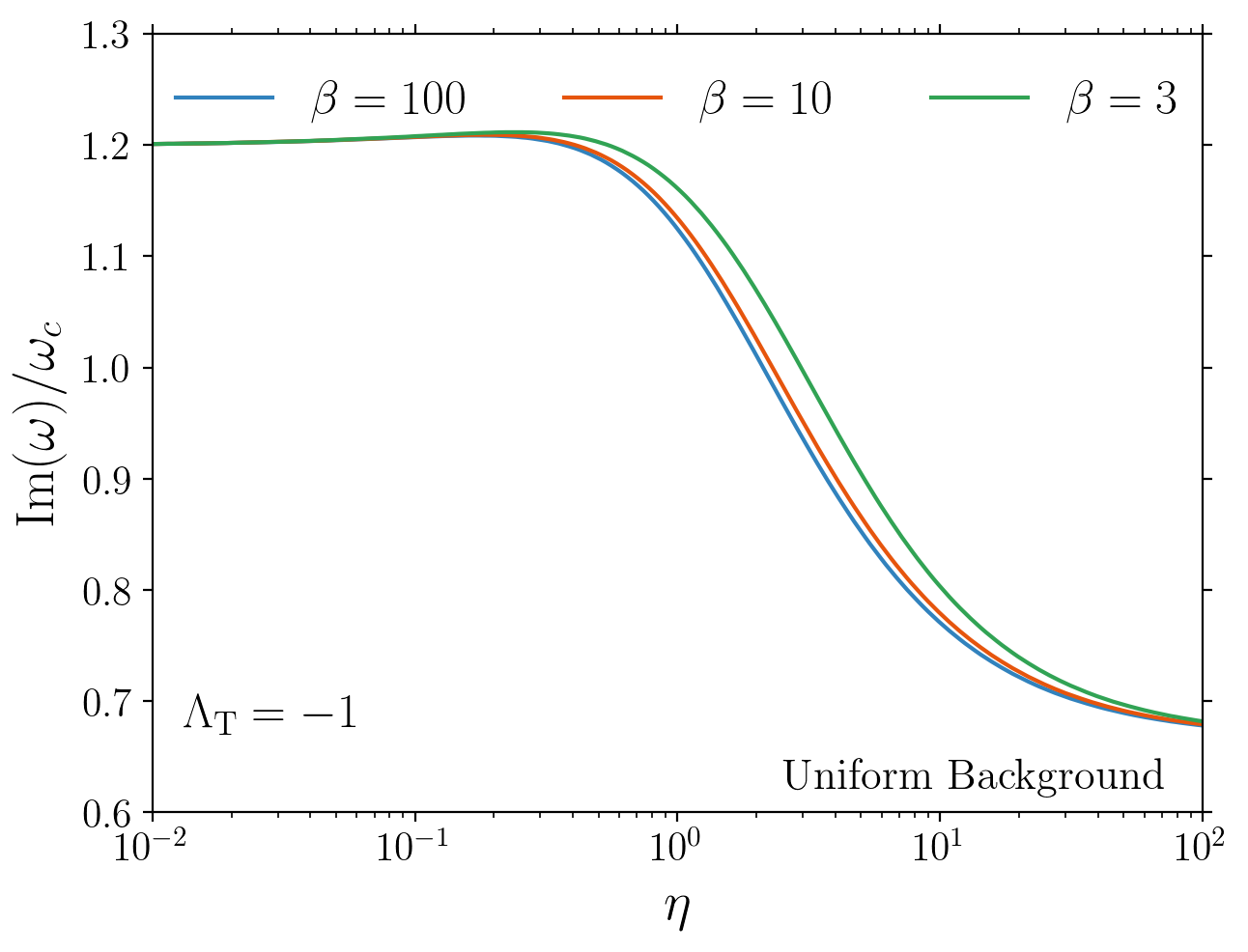}
  \end{minipage}
    \begin{minipage}[b]{0.32\textwidth}
    \includegraphics[width=\textwidth]{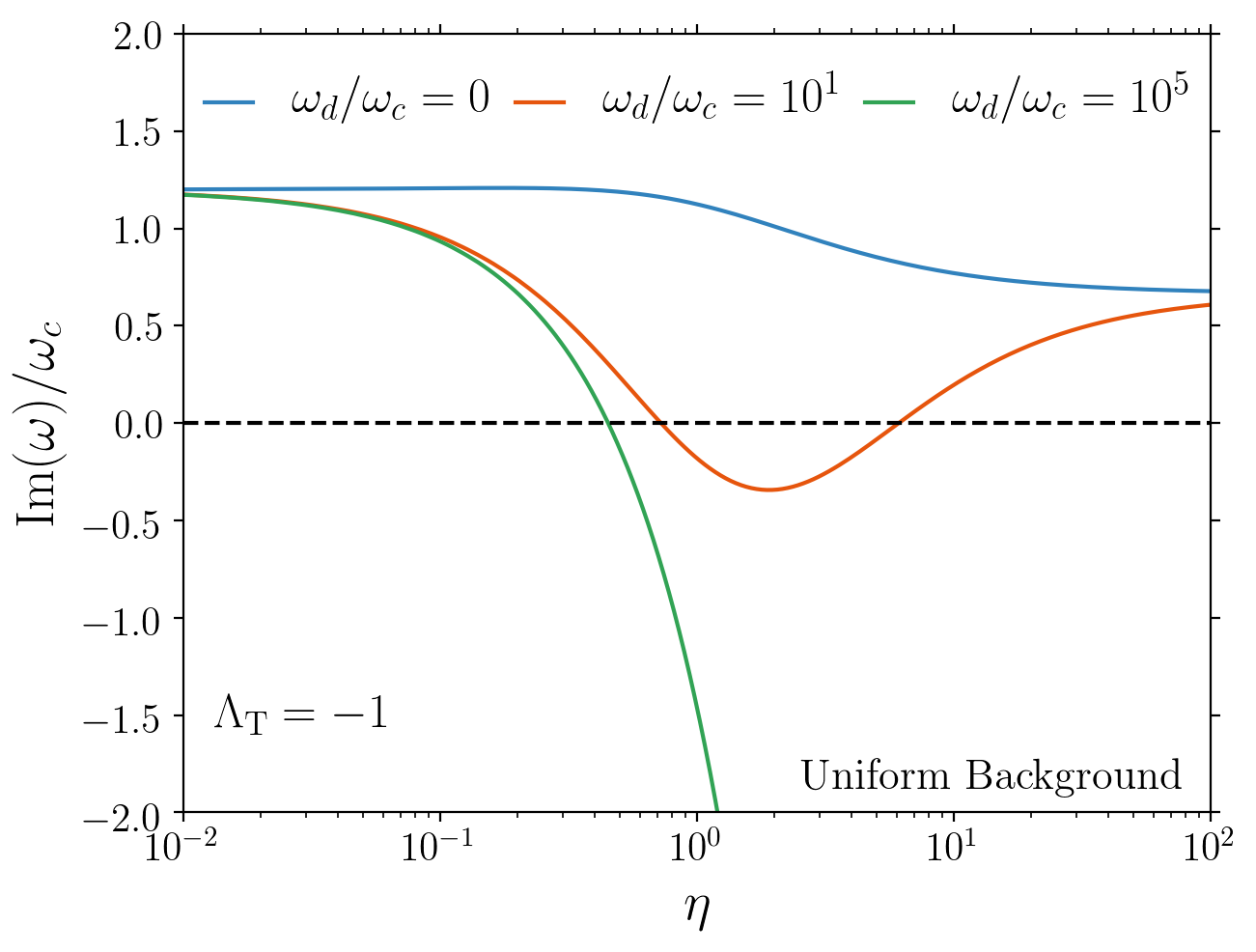}
  \end{minipage}
    \begin{minipage}[b]{0.32\textwidth}
    \includegraphics[width=\textwidth]{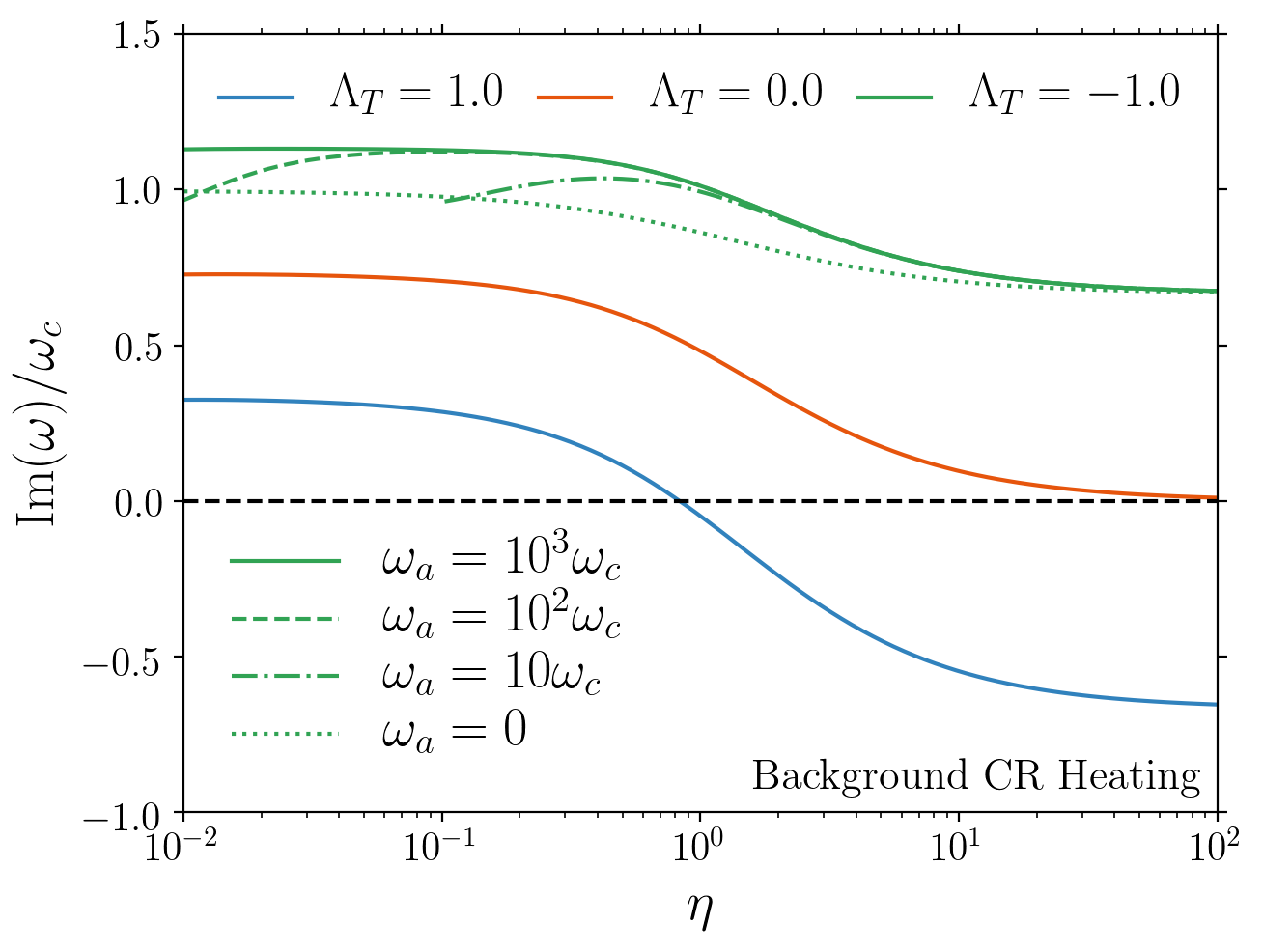}
  \end{minipage}
  \begin{minipage}[b]{0.32\textwidth}
    \includegraphics[width=\textwidth]{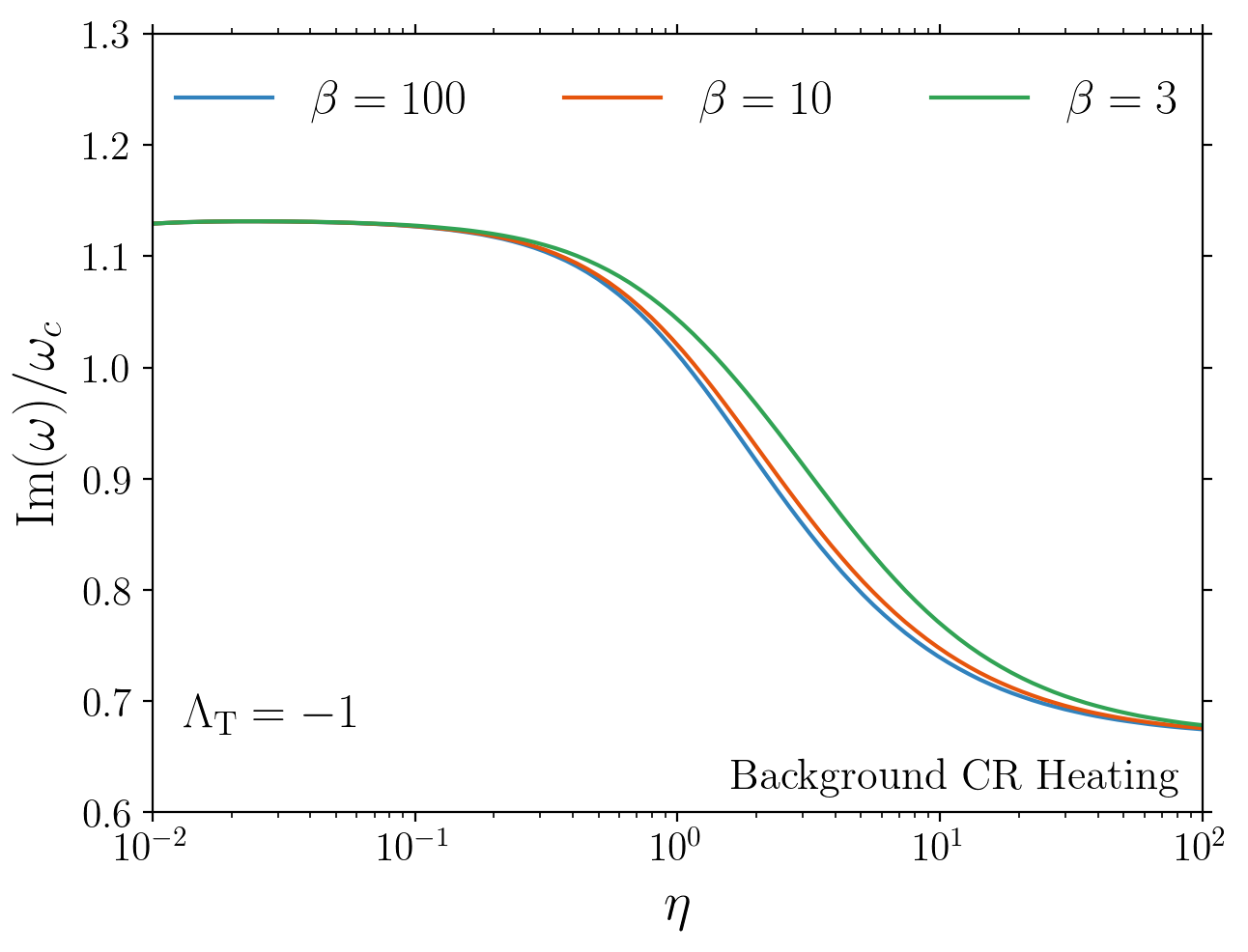}
  \end{minipage}
    \begin{minipage}[b]{0.32\textwidth}
    \includegraphics[width=\textwidth]{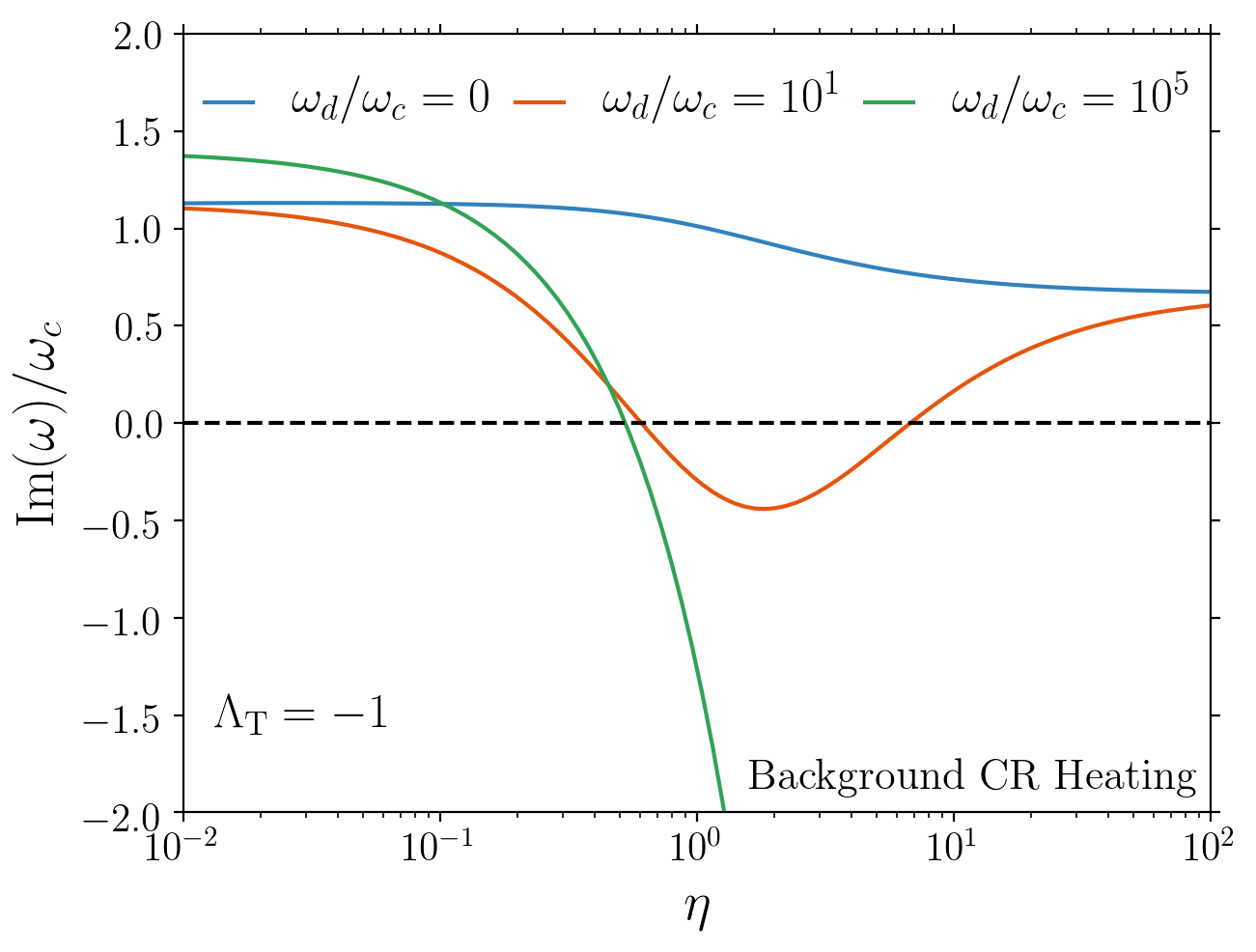}
  \end{minipage}

  \caption{Thermal instability growth rates as a function of $\eta$. ${\rm Im} (\omega) > 0$ corresponds to growing modes. Unless explicitly stated otherwise in the plots, the presented growth rates are for our fiducial parameters ($\omega_a = 10^3 \omega_c$ and $\beta =100$). We consider smaller $\omega_a$ in the left panels (with $\beta = 100$ fixed) and smaller $\beta$ in the middle panels  (with $\omega_a = 10^3 \omega_c$ fixed).  \textbf{Top panels:} Thermal instability in uniform medium. \textbf{Left:} growth/damping rate for different cooling curve slopes. \textbf{Middle:} $\LT = -1$ growth rate for different $\beta$. For $\beta \gtrsim 3$, the high-$\beta$ result is a very good approximation. \textbf{Right:} impact of diffusion on modes with $\bm{k\cdot B} \neq 0$. \textbf{Bottom panels:} same as top panels, but for a background with cosmic-ray heating balancing cooling, but no gravity. The small-$\eta$ limit is different for perpendicular modes with $\omega_a = 0$ (dotted green line in left panel; see Section \ref{sec:perp_modes}). For $\omega_a = 10 \omega_c$ in the left panel (dash-dotted green line), we only plot the growth rate for $\eta > 0.1$, where our WKB analysis is applicable. In the right plot, the plateau at small $\eta$  depends on whether diffusion is more important than streaming (i.e. whether $\omega_d > \omega_a$ or $\omega_a > \omega_d$). \label{fig:uni}} 
\end{figure*}

We now look back at the dispersion relation in \eqref{eq:dispersion_quad_uni}. How the solutions of  \eqref{eq:dispersion_quad_uni} depend on $\eta$ is particularly transparent. We can read off the solutions in the limits $\eta \rightarrow 0$ and $\eta \rightarrow \infty$. \footnote{Note that in our notation the limits $\eta \rightarrow 0$ ($\eta \rightarrow \infty$) mean that $\eta$ is much smaller (larger) than any other dimensionless parameter in the problem, e.g. $\omega_d / \omega_a$, $\omega_a / \omega_c$ etc.} As $\eta \rightarrow 0$, the unstable gas entropy mode is just the standard isobaric thermal instability result, with a small oscillatory part due to the perturbed CR heating:\footnote{The real  (oscillatory) part of the solution in eq. \ref{eq:small_eta_uni} also assumes $\omega_a > \omega_d$ (for $\omega_d > \omega_a$ the real part vanishes as $\delta p_c$ is suppressed by diffusion).}
\begin{equation} \label{eq:small_eta_uni}
    \omega = -\frac{4}{15}\eta \omega_a + \frac{2}{5}i \Big( 2 - \LT  \Big) \omega_c,
\end{equation}
which comes from the isobarically perturbed cooling function, $\delta (- \rho^2 \Lambda)=-\omega_c p_g \ (2 - \LT)  \delta \rho / \rho$ (at small $\eta$ we have that $\delta p_g / p_g \ll \delta \rho / \rho$, as discussed in Section \ref{sec:perts}).  As $\eta \rightarrow \infty$, we get an overstable solution:
 \begin{equation} \label{eq:etainfy_uni}
     \omega = - \frac{2}{3} \omega_a - \frac{2}{3} i  \LT \omega_c.
 \end{equation}
 Note that the $-(2/3) \omega_a$ comes from the perturbed CR heating, as discussed in \ref{sec:uni_streaming}, while the $-(2/3)\LT \omega_c$ growth rate comes from the isochorically perturbed cooling function (recall that in the limit $\eta \rightarrow \infty$, $\delta p_g / p_g \gg \delta \rho / \rho$, so that unstable modes are isochoric). CR heating does not directly affect the growth rate. Equations \ref{eq:small_eta_uni} and \ref{eq:etainfy_uni} are consistent with the low and high $\eta$ limits in Figure \ref{fig:uni} (upper panels).
 
\subsection{Effect of CR Diffusion} \label{sec:uni_diff}
CR diffusion does not suppress the overall excitation of thermal instability. It nevertheless suppresses the growth of some modes which would otherwise be thermally unstable (see, e.g., top right panel of Figure \ref{fig:uni}). 

\begin{figure*}
  \centering
  \begin{minipage}[b]{0.45\textwidth}
    \includegraphics[width=\textwidth]{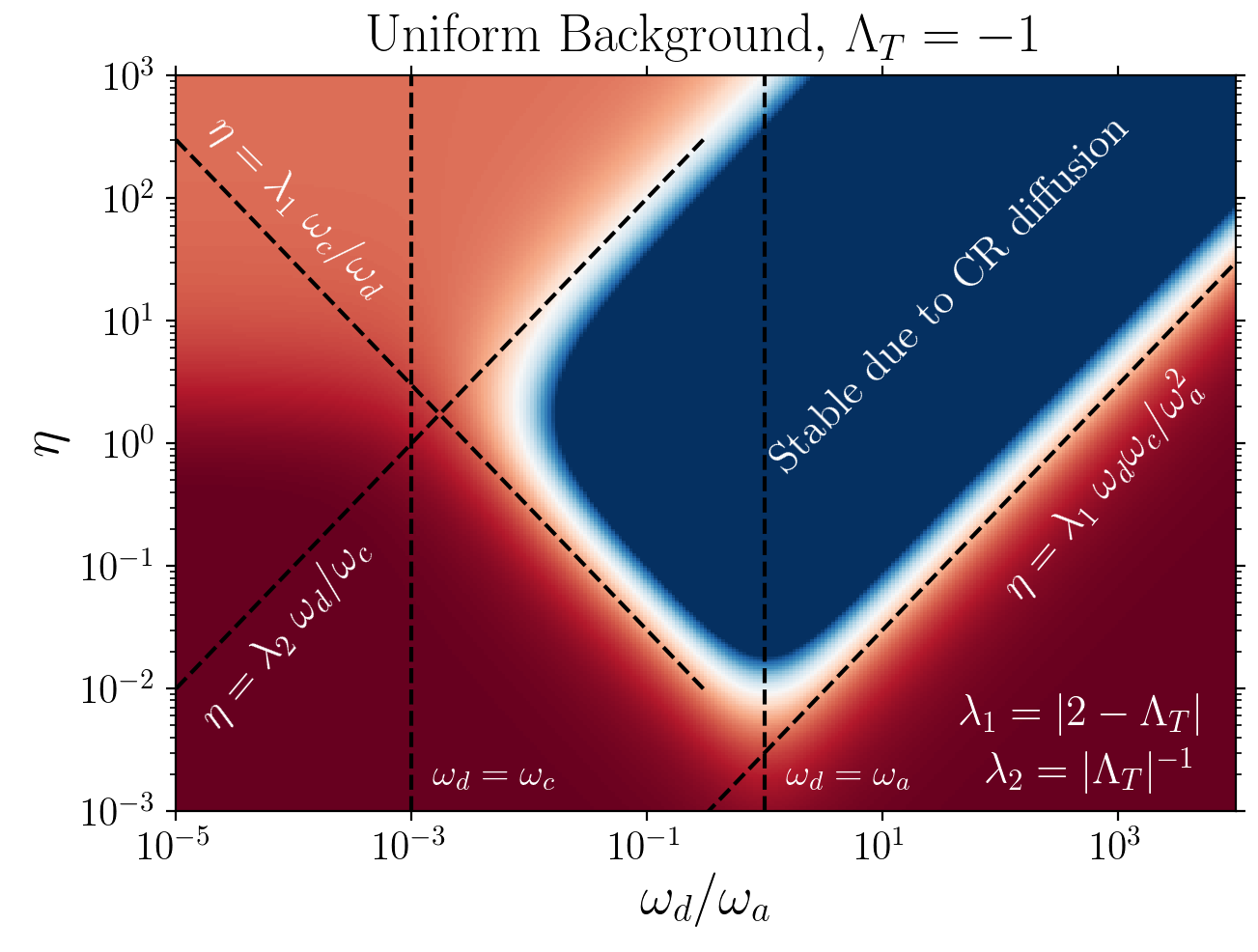}
  \end{minipage}
  \begin{minipage}[b]{0.45\textwidth}
    \includegraphics[width=\textwidth]{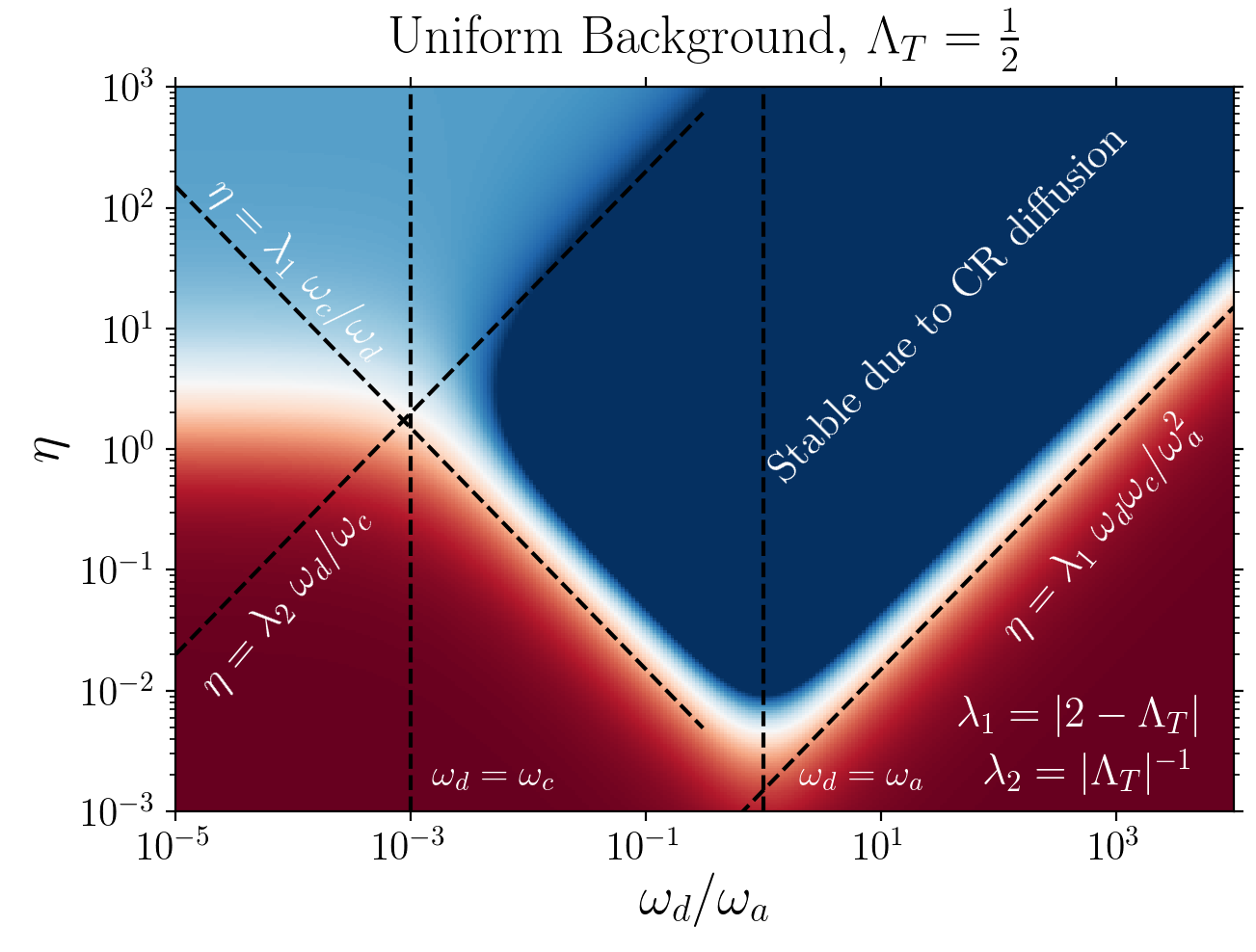}
  \end{minipage}
  \begin{minipage}[b]{0.45\textwidth}
    \includegraphics[width=\textwidth]{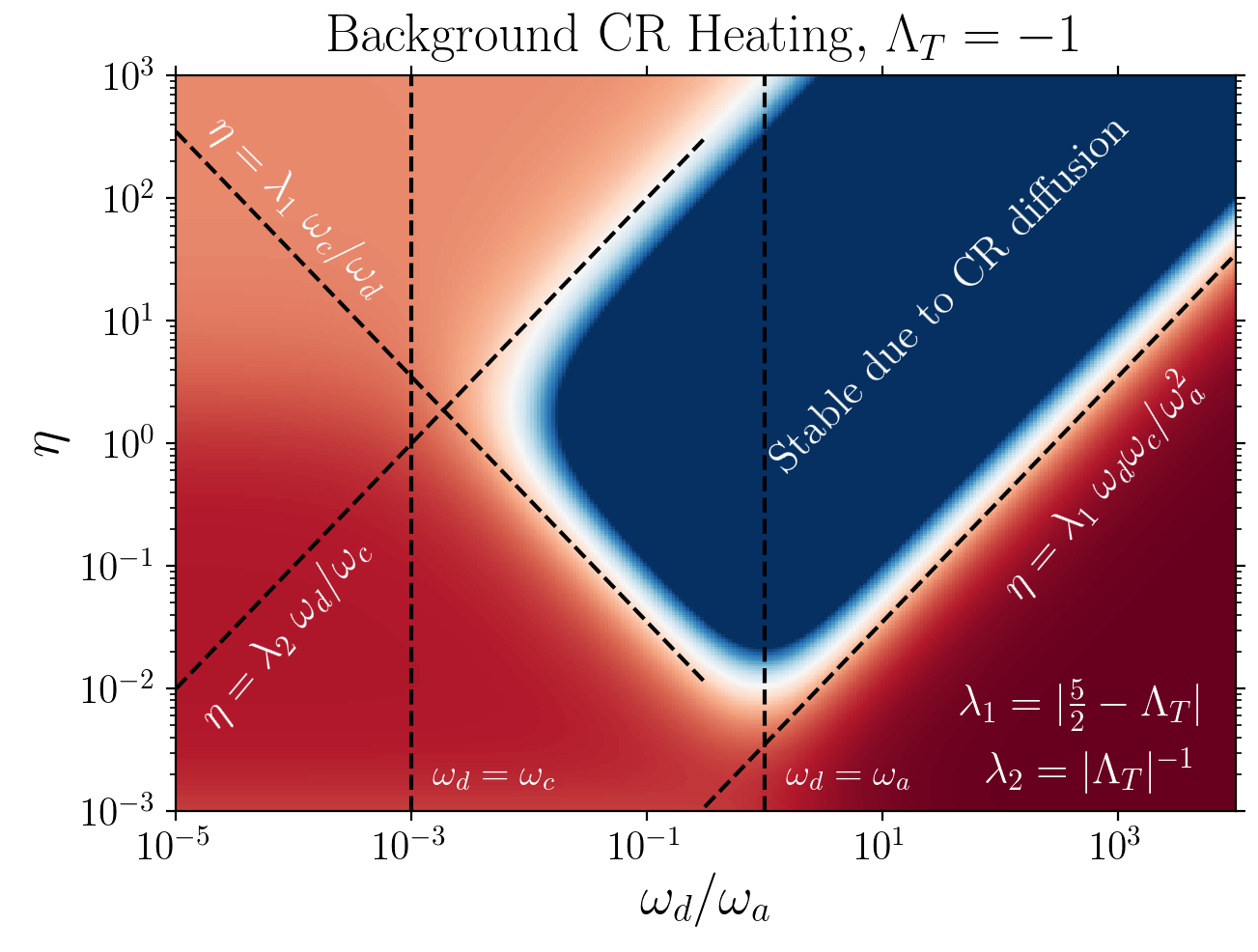}
  \end{minipage}
  \begin{minipage}[b]{0.45\textwidth}
    \includegraphics[width=\textwidth]{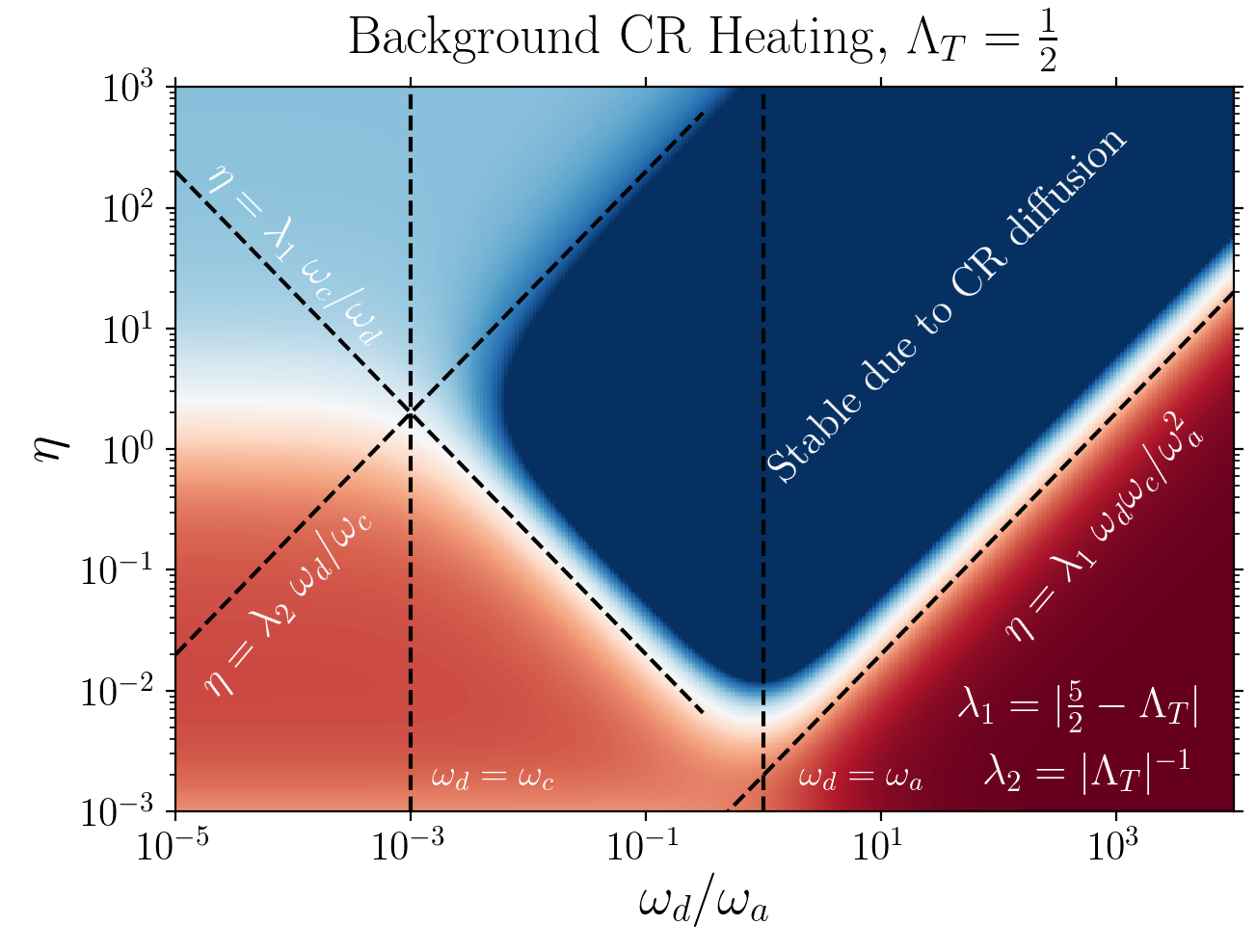}
  \end{minipage}
  \caption{Effect of CR diffusion on thermal instability. We show thermal stability/instability boundaries of modes with $\bm{k \cdot B} \neq 0$ in the ($\eta$, $\omega_d / \omega_a$) plane, for  $\omega_a = 10^3 \omega_c$ and $\beta \rightarrow \infty$ (the fiducial $\beta = 100$ case looks the same). ${\rm Im} (\omega) > 0$ (growing modes) are shown in red, ${\rm Im} (\omega) < 0$ (decaying modes) are shown in blue. \textbf{Top panels:} Stability/instability boundaries in uniform medium. Left: $\LT = -1$. Right: $\LT = 1/2$. The dark blue shows the region where thermal instability is suppressed by CR diffusion. The light blue shows thermal stability due to $\LT > 0$ (eq. \ref{eq:etainfy}). The approximate boundaries (dashed lines) of the diffusion-affected region are derived in Appendix \ref{app_diff}. \textbf{Bottom panels:} same as top panels, but in a background with CR heating balancing cooling (but no gravity). Note that the growth rate at small $\eta$ (dark vs light red) now depends on whether $\omega_a > \omega_d$ or $\omega_d > \omega_a$ ($\rm{Im}(\omega) = (2/5) \omega_c \Big( 11/6 - \LT \Big) $ and $\rm{Im}(\omega) = (2/5) \omega_c \Big( 5/2 - \LT \Big)$ respectively).  \label{fig:uni_map}} 
\end{figure*}

We can study the effects of CR diffusion on modes with $\bm{k \cdot B} \neq 0$ by looking at thermal stability maps in the $(\eta, \omega_d / \omega_a)$ plane. We show this in Figure \ref{fig:uni_map}. The results of the uniform medium calculation are shown in the top panels, for $\LT < 0$ (left panel, $\LT = -1$) and for $2 > \LT> 0$ (right panel, $\LT = 1/2$). The blue color corresponds to stable solutions, red denotes growing (i.e. thermally unstable) solutions. We provide approximate boundaries for the region of parameter space where cosmic-ray diffusion can suppress thermal instability (dashed lines). A heuristic derivation of these boundaries can be found in Appendix \ref{app_diff}.

Here we summarise the main results from Appendix \ref{app_diff}. For modes with $\omega_d < \omega_a$, diffusion suppresses thermal instability if $\eta$ satisfies:
\begin{equation} \label{eq:weakdiff_cond}
   |2 - \LT | \frac{\omega_c}{\omega_d}  \lesssim  \eta \lesssim | \LT |^{-1} \frac{\omega_d}{\omega_c} .
\end{equation}
If  $\eta$ is too small for the above condition to be satisfied, the instability is isobaric, with ${\rm Im}(\omega) = (2/5)  \omega_c \Big(2 - \LT \Big)$. If $\eta > | \LT |^{-1} \omega_d / \omega_c $, the growth rate approaches the asymptotic limit ${\rm Im}(\omega) = - (2/3) \LT \omega_c $ from equation \ref{eq:etainfy_uni}.

For modes with $\omega_d > \omega_a$, diffusion suppresses thermal instability if $\eta$ satisfies:
\begin{equation} \label{eq:strongdiff_cond}
   |2 - \LT | \frac{\omega_d \omega_c}{\omega_a^2}  \lesssim  \eta \lesssim  | \LT  |^{-1} \frac{\omega_d}{\omega_c} .
\end{equation}
If  $\eta <  |2 - \LT | \omega_d \omega_c / \omega_a^2  = |2 - \LT | \kappa \omega_c / v_A^2$, the instability is again isobaric, with ${\rm Im}(\omega) = (2/5) \Big(2 - \LT\Big) \omega_c$. When $\eta$ is large, the growth rate again approaches the asymptotic limit ${\rm Im}(\omega) = - (2/3) \LT \omega_c $. Note that in the limit $\kappa \rightarrow \infty$, the instability is isobaric for arbitrary $\eta$ because CR diffusion suppresses $\delta p_c$ (recall that in the high-$\beta$ limit we have that $\delta p_c \approx - \delta p_g$, so $\delta p_g \approx 0$ if $\delta p_c$ is suppressed by CR diffusion).

\subsubsection{Cosmic-Ray Field Length} \label{sec:CR_field_length}
In Appendix \ref{app_diff} we show that CR diffusion can suppress thermal instability because it affects the thermal gas in a way akin to thermal conduction (mediated by the perturbed CR heating term, see Appendix \ref{sec:app_weakdif}). This suggests that there is a CR-diffusion analogue of the Field length for thermal conduction (\citealt{field65}), below which thermal instability is suppressed.

In Appendix \ref{app:crdiff_field} we show that the dimensionless ratio $\kappa \omega_c / ( \eta v_A^2)$, the ratio of the cooling rate to the CR-heating rate at high-$k$, determines whether there is a Field length associated with CR diffusion. If $\kappa \omega_c / ( \eta v_A^2) \gtrsim 1$  then CR diffusion does not suppress thermal instability of high-$k$ modes ($\omega_d \gg \omega_a$), as the cooling rate exceeds the CR heating rate. There is no ``CR Field length" below which thermal instability is completely suppressed. Instead, the instability of high-$k$ modes is isobaric with growth rates ${\rm Im}(\omega) = (2/5) \Big(2 - \LT\Big) \omega_c$ (as $\delta p_g \approx - \delta p_c$ is suppressed by CR diffusion). On the other hand, if $\kappa \lesssim \eta v_A^2 / \omega_c$, there is a maximum $\bm{\hat{b} \cdot k}$ at which thermal instability occurs (Figure \ref{fig:diff_vs_k}). The CR Field length is (Appendix \ref{app:crdiff_field}):
\begin{equation} \label{eq:cr_field}
  \lambda_{\rm CRF} \sim
    \begin{cases}
      2 \pi |\bm{\hat{b} \cdot \hat{k}}|  \sqrt{\frac{\eta \kappa}{\omega_c}} & \eta < 1\\
            2 \pi |\bm{\hat{b} \cdot \hat{k}}|  \sqrt{\frac{ \kappa}{\eta \omega_c}} & \eta > 1.
    \end{cases}       
\end{equation}

Note that the CR Field length is very similar to the classic Field length with the thermal diffusion coefficient replaced by the CR diffusion coefficient. We can estimate  $\kappa \omega_c / \eta v_A^2$ for CGMs of Milky-Way-like galaxies:
\begin{equation}
    \frac{\kappa \omega_c}{\eta v_A^2} \sim 1 \  \frac{\kappa}{10^{28} \ {\rm cm^2 \ s^{-1}}} \frac{\omega_c}{ 10^{-15} \ {\rm s^{-1}}}  \Big( \frac{\eta}{1} \Big)^{-1} \Big( \frac{v_A}{3 \times 10^6 \ {\rm cm \ s^{-1}}} \Big)^{-2}.
\end{equation}
We chose $\kappa = 10^{28} \ {\rm cm^2 \ s^{-1}}$ motivated by diffusion-only models of CR observations in the Milky Way, which infer $\kappa \sim 10^{28} - 10^{29} \ {\rm cm^2 \ s^{-1}}$ depending on the size of the CR halo (e.g., \citealt{Linden2010}). It is plausible that $\kappa \omega_c / \eta v_A^2 > 1$, so that CR diffusion does not suppress thermal instability at small scales. However, if instead $\kappa \omega_c / \eta v_A^2 < 1$ (e.g., if streaming is the dominant transport process $\kappa$ may be $\ll 10^{28} \ {\rm cm^2 \ s^{-1}}$), thermal instability of modes with wavelengths smaller than the CR Field length,
\begin{equation}
    \lambda_{\rm CRF}  \sim 7 \ {\rm kpc} \ |\bm{\hat{b} \cdot \hat{k}}| \ \Big(\frac{\kappa}{10^{28} \ {\rm cm^2 \ s^{-1}}} \Big)^{1/2} \Big( \frac{\omega_c}{ 10^{-15} \ {\rm s^{-1}}},  \Big)^{-1/2}
\end{equation}
is suppressed by CR diffusion (here we assumed $\eta \sim 1$).

\subsection{Thermal Stability versus Instability} \label{sec:inst_cond}

 \begin{figure}
  \centering
    \begin{minipage}[b]{0.45\textwidth}
    \includegraphics[width=\textwidth]{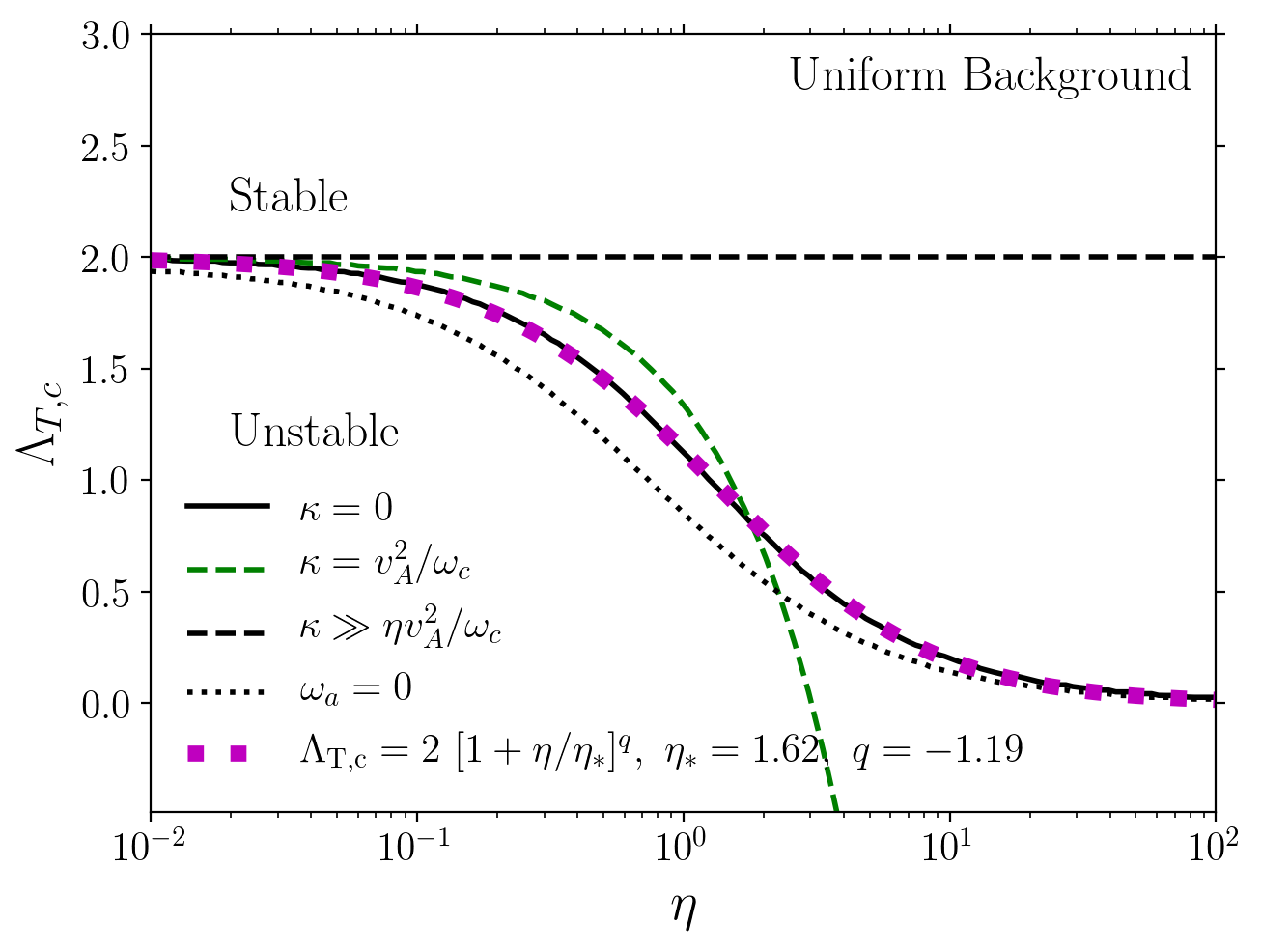}
  \end{minipage}
     \begin{minipage}[b]{0.45\textwidth}
    \includegraphics[width=\textwidth]{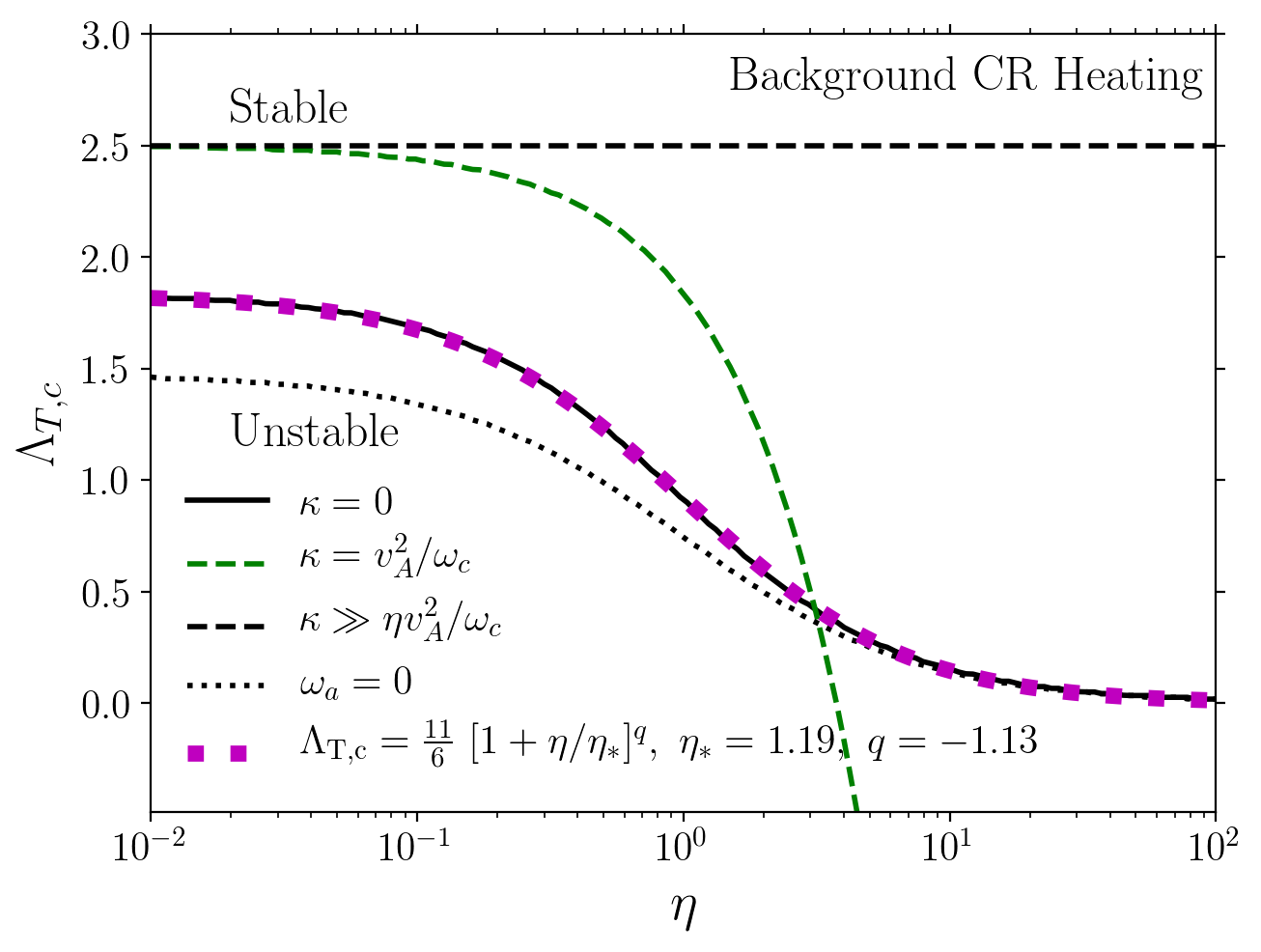}
  \end{minipage}
  \caption{$\LTC$ versus $\eta \equiv p_c / p_g$, where $\LTC$ is the $\partial \ln \Lambda / \partial \ln T$ that defines the boundary between overall thermal stability and instability. For a given $\eta$, thermal instability occurs if $\LT < \LTC$.  \textbf{Top:} $\LTC$ in a uniform medium. \textbf{Bottom:} $\LTC$ in a medium with background CR heating. We use $\beta =100$ and we include modes that satisfy $\omega_a > 10 \omega_c$ ($\omega_a > 10 \omega_c$ and $\omega_a > 10 \omega_c \eta^{-1}$) in the uniform (CR-heated) background. $\LTC$ does  not change significantly for $\beta \gtrsim 3$, and for $\omega_a \gtrsim \omega_c$ ($\omega_a \gtrsim \omega_c \eta^{-1}$) in the uniform (CR-heated) background absent diffusion (i.e. growth rates are approximately constant for local perturbations satisfying $kH \gtrsim 1$; in Figure \ref{fig:uni} we show how growth rates depend on $\omega_a$ and $\beta$). The solid black line is for $\kappa = 0$ (no CR diffusion) and the dotted magenta line is a simple broken power-law fit. The horizontal dashed line is the stability/instability boundary when CR diffusion is present and $\kappa \gg \eta v_A^2 / \omega_c$, while the green dashed line shows the boundary for $\kappa = v_A^2 / \omega_c$. The dotted lines are the thermal instability boundaries for perpendicular modes only (i.e. $\omega_a = \omega_d = 0$). \label{fig:uni_boundary} }
\end{figure}

In addition to the slope of the cooling function, $\LT$, thermal stability clearly also depends on the CR pressure fraction, $\eta$ (which sets whether perturbations are isobaric or isochoric, see Section \ref{sec:perts}). We show the ``critical" cooling function logarithmic slope, $\LTC$, demarcating the boundary between thermal stability and instability to any local perturbation (satisfying $\omega_a \gg \omega_c$), in the top panel of Figure \ref{fig:uni_boundary}. The solid black line shows the boundary without CR diffusion, i.e. $\kappa=0$, and the magenta dotted line is a simple broken power-law fit of the form
\begin{equation}
	\LTC = 2 [ 1+ \eta / \eta_*  ]^q,
\end{equation}
where $\eta_* = 1.62$ and $q = -1.19$ are the best-fit parameters. For a given $\eta$, the system is thermally unstable if $\LT < \LTC$. The dotted line is the thermal stability boundary for perpendicular modes only, i.e. for $\omega_a = 0$, for which $p_c$ and $\rho$ follow an adiabatic relation with index $4/3$ (note that our thermal stability criterion is not the same as in \citealt{pfrommer13}; see last paragraph of Section \ref{sec:uni_disp}). It is notable that the $\omega_a = 0$ and $\omega_a \gg \omega_c$ values of $\LTC$ in Figure \ref{fig:uni_boundary} are very similar. This again highlights that perturbed CR heating does not significantly affect the growth rates of thermal instability. Instead, it turns a purely growing mode into an overstability (see Section \ref{sec:uni_streaming}).

The dashed horizontal line $\LTC = 2$ is the thermal stability/instability boundary if CR diffusion is present and $\kappa \gg \eta v_A^2 / \omega_c$. $\LTC = 2$ due to the fact that for $\kappa \gg \eta v_A^2 / \omega_c$ high-$k$ perturbations are isobaric, and so high-$k$ modes (with $\bm{k \cdot B} \neq 0$) always have a growth rate ${\rm Im}(\omega) = (2/5)\omega_c \Big( 2 - \LT  \Big)$ (see discussion in Section \ref{sec:CR_field_length}). For $\kappa \lesssim \eta v_A^2 / \omega_c$, high-$k$ perturbations are suppressed by CR diffusion. Only modes with wavelengths longer than the CR Field length can be thermally unstable. For $\kappa = v_A^2 / \omega_c$ (green dashed line) and $\eta > 1$, the CR Field length is at lower $k$ than the modes used for the stability boundary calculation ($\omega_a > 10 \omega_c$) and so CR diffusion suppresses thermal instability of these modes. Note that for $\kappa = v_A^2 / \omega_c$ and $\eta \ll 1$ $\LTC = 2$, as high-$k$ modes are isobaric at low $\eta$.  We stress again that perpendicular modes ($\bm{k \cdot B = 0}$, dotted line) are not affected by CR diffusion.

\subsection{Photoionization Equilibrium} \label{sec:pie}
We can easily extend our CR thermal instability analysis to a background in photoionization equilibrium (PIE), with no background CR heating, but where CR heating is still present in the perturbed equations. We can then treat PIE analogously to our uniform background, by absorbing photoionization heating and cooling into an effective cooling function $\Lambda$. In PIE, this effective cooling function satisfies $\LT > 2$ (e.g., \citealt{wiersma09}), and so such systems are thermally stable for any CR pressure fraction $\eta$.

\section{Equilibrium with cosmic-ray heating balancing cooling} \label{sec:nonuni}
\subsection{Equilibrium} \label{sec:nonuni_eq}
In this section we look at equilibria in which cooling is completely balanced by cosmic-ray heating,
\begin{equation} 
    - \bm{v_A \cdot \nabla}p_c =  \rho^2 \Lambda(T) =  \omega_c p_g.
\end{equation}
We still ignore gravity, i.e. we set $\bm{g} = 0$. Hydrostatic equilibrium implies that the CR pressure gradient is balanced by the gas pressure gradient:
\begin{equation}
    \bm{\nabla}p_c = -\bm{\nabla}p_g.
\end{equation}
Without loss of generality, we assume that the variation is purely in the vertical direction, i.e. $\bm{\nabla}p_c = (\partial p_c / \partial z) \ez$. We assume a uniform magnetic field, $\bm{B} = B \sin \theta_B \ex$ + $B \cos \theta_B \ez$.

We choose the background pressures such that they have a linear profile, i.e. $\partial p_c / \partial z =  {\rm const}$, so that CR diffusion does not enter in the equilibrium setup. The cosmic-ray pressure equation \eqref{eq:pc} then implies that  $p_c  \propto \rho^{2/3}$ and $\bm{\nabla \cdot v_A} = (3/4) \omega_c \eta^{-1}$.

\subsection{Linear Perturbations in 1 Dimension} \label{sec:nonuni_1d}
The background  gradients give rise to extra terms in linear perturbation theory, which modify equations \ref{eq:uni_drho}--\ref{eq:uni_pc}. We show the linearised equations in a medium with background CR heating in Appendix \ref{app:nonuni}, which we again solve using MATLAB (there is again little physical insight gained from explicitly writing down the 6th-order dispersion relation). 

In addition to explicitly solving equations \ref{eq:nonuni_drho}--\ref{eq:nonuni_pc}, we consider the simpler 1-dimensional problem in which $\bm{B}$, $\bm{k}$, $\bm{\xi}$ (as well as the background gradients) are  all along $\bm{\hat{z}}$. This is motivated by the fact that we find that 1D thermal instability growth rates agree essentially perfectly with the more general calculation with $\bm{B}$, $\bm{k}$ and $\bm{\hat{z}}$ not aligned (unless $\bm{k \cdot B}$ = 0, which we treat separately in Section \ref{sec:perp_modes}). We explain why the 1D calculation correctly predicts the gas entropy mode eigenfrequency in Appendix \ref{app:1D_val}.  In the high-$\beta$ limit, the 1D dispersion relation simplifies to a quadratic:

\begin{gather}
\begin{aligned} \label{eq:disp_1d}
   0 &=  \Big( \frac{4}{3} \eta \omega -  \frac{2}{3} \eta \omega_a + i \omega_c     - \omega \omega_c  \big(\frac{3}{2} \omega_c \eta^{-1} -i\omega_a \big)^{-1}    \Big) \\ & \Big( \frac{3}{2}  \omega   +  \omega_a  + i \omega_c \LT \Big) \   + \  \Big(  \omega - \omega_a + i \omega_d  +   i \omega_c \eta^{-1}  \Big)  \\ & \Big(\frac{5}{2}  \omega  - i \omega_c \big(\frac{5}{2} - \LT \big) - \frac{3}{2}\omega \omega_c  \big( 1+\frac{5}{2}\eta^{-1} \big) \big( \frac{3}{2} \omega_c \eta^{-1} - i \omega_a \big)^{-1} \Big).  
\end{aligned}
\end{gather}
Note that $\omega_a$ again shows up due to the perturbed CR heating, and not as a result of the perturbed magnetic field (indeed, $\bm{\delta B} =0$ in 1D).

Thermal instability growth rates as a function of $\eta$ in a medium with background CR heating are shown in the bottom panels of Figure \ref{fig:uni}. Unless explicitly stated otherwise, the growth rates are for our fiducial parameters, $\omega_a = 10^3 \omega_c$ and $\beta = 100$. The $\beta \rightarrow \infty$ growth rates calculated from \eqref{eq:disp_1d} overlap almost perfectly with the $\beta = 100$ calculation (and are therefore not explicitly plotted). The bottom left panel shows the growth rates for different cooling curve slopes $\LT$. We also show how $\LT=-1$ growth rates change for smaller $\omega_a$: $\omega_a = 10^2 \omega_c$, $\omega_a = 10 \omega_c$ and $\omega_a = 0$ ($\bm{k \cdot B }= 0$, see Section \ref{sec:perp_modes}). The middle panel shows how the $\beta = 100$ growth rate ($\approx$ $\beta \rightarrow \infty$ growth rate)  compares to the growth rate at smaller $\beta$ and the same $\LT$. The right panel shows the effects of diffusion (again for $\LT = -1.0$), in the limits $\omega_d = 0$ (blue), $\omega_a \gg \omega_d \gg \omega_c$ (orange) and $\omega_d \gg \omega_a$ (green). Note that unlike the uniform-medium calculation, the growth rate at small $\eta$ now depends on whether the mode is diffusion dominated (i.e. whether $\omega_d > \omega_a$, see \ref{sec:nonuni_asympt}).

\subsubsection{Asymptotic Limits} \label{sec:nonuni_asympt}
We now  consider the asymptotic limits in the presence of cooling, CR streaming and diffusion. The $\eta \rightarrow \infty$ limit is again simple and can be read off directly from \eqref{eq:disp_1d}. The solution is overstable,
 \begin{equation} \label{eq:etainfy}
     \omega = - \frac{2}{3} \omega_a - \frac{2}{3} i  \LT \omega_c ,
 \end{equation}
 and is identical to  the uniform-medium large-$\eta$ result.  The $-(2/3) \omega_a$ oscillation frequency again comes from the perturbed CR heating (see Section \ref{sec:uni_streaming}), which turns thermally unstable modes into propagating waves. 

The small-$\eta$ limit ($\omega_a \gg \omega_c \eta^{-1} \gg \omega_c$)\footnote{Recall that in our local analysis we only consider  $\omega_a > \omega_c \eta^{-1}$. This corresponds to perturbations that satisfy $kH \gtrsim 1$, $H$ being a characteristic background length scale.} depends on whether the mode is streaming or diffusion dominated. In the streaming-dominated case (i.e. modes with $\omega_d \ll \omega_a$), the isobaric growth rate, $(2/5) \omega_c (2 - \LT)$, which comes from isobaric perturbations to the cooling function, $\delta (-\rho^2 \Lambda)=-\omega_c p_g  \ (2 - \LT) \delta \rho / \rho$, is modified by CR streaming and background heating,\footnote{This can be shown by solving eq. \ref{eq:disp_1d} perturbatively using the ordering $\omega_a \gg \omega_c \eta^{-1} \gg \omega_c$.}
\begin{equation} \label{eq:sol_116}
    \omega = - \frac{4}{15}\eta \omega_a + i \frac{2}{5} \omega_c \Big( \frac{11}{6} - \LT \Big).
\end{equation}
 For modes with  $\omega_d \gg \omega_a$, one can show that 
\begin{equation} \label{eq:small_eta_nonuni}
    \omega = i \frac{2}{5}\omega_c \Big(\frac{5}{2} - \LT \Big).
\end{equation}
In this strong-diffusion limit, the $5/2$ (instead of 2) arises from the perturbed CR heating term, $- \bm{\delta v_A \cdot \nabla} p_c = - (1/2) \omega_c p_g \delta \rho/\rho$. Note that diffusion suppresses CR pressure perturbations, so that $-\bm{v_A \cdot \nabla} \delta p_c$ is suppressed and does not give rise to gas-entropy oscillations (i.e., the mode is purely growing, unlike eq. \ref{eq:sol_116}).

These results differ modestly from the uniform medium calculation (see Section \ref{sec:uni_asymptotes} or compare the top and bottom panels of Figure \ref{fig:uni}), as now the background CR pressure gradient modifies the growth rate.

\subsection{Perpendicular Modes}
\label{sec:perp_modes}
\subsubsection{Dispersion Relation}
The 1D calculation does not apply to modes propagating perpendicular to the magnetic field direction, such that $\bm{k \cdot B} = 0$. In this case, we can obtain an approximate quadratic dispersion relation by taking the high-$\beta$ limit and dropping advective ($\bm{\xi \cdot \nabla}$) background-gradient terms:
\begin{equation} \label{eq:disp_perp}
    \Big( \omega + i \omega_c \eta^{-1} \Big) \Big( \frac{5}{2}\omega - i \Big( \frac{3}{2} - \LT \Big)\omega_c \Big)
    +  \Big( \frac{4}{3} \eta \omega + i \omega_c \Big) \Big( \frac{3}{2}\omega +i \omega_c \LT   \Big) = 0.
\end{equation}
Note that $\omega_a = \omega_d = 0$ for modes with $\bm{k \cdot B} = 0$, and  so do not show up in the above dispersion relation.

\subsubsection{Asymptotic Limits}

The $\eta \rightarrow \infty$ growth rate does not change, and the solution is now a purely growing mode (as $\omega_a = 0$):
 \begin{equation} \label{eq:etainfy_perp}
     \omega = - \frac{2}{3} i  \LT \omega_c .
 \end{equation}
For  $\eta \rightarrow 0$, the mode is also purely growing, with
 \begin{equation} \label{eq:etasmall_perp}
     \omega =  \frac{2}{5} i \Big(\frac{3}{2}-  \LT \Big) \omega_c.
 \end{equation}
This differs from the corresponding limit in Section \ref{sec:nonuni_asympt}, as for perpendicular modes the perturbed CR heating is  $- \bm{\delta v_A \cdot \nabla} p_c \approx  (1/2) \omega_c p_g \delta \rho / \rho$, while the isobarically perturbed cooling function is still $-\omega_c p_g \ (2 - \LT)  \delta \rho / \rho$. Note that because $\bm{k \cdot B} =0$, there are no entropy oscillations driven by CR heating. The growth rate of perpendicular modes as a function of $\eta$ (for $\LT=-1$) is plotted as a dotted green line in the lower-left panel of Figure \ref{fig:uni}.

\subsection{Effect of CR Diffusion}
The bottom panels of Figure \ref{fig:uni_map} show stability maps of modes with $\bm{k \cdot B} \neq 0$ in the $\Big( \eta, \omega_d / \omega_a \Big)$ plane in a medium with background CR heating. Once again, blue denotes stable solutions, while red denotes growing solutions. The left panel is for $\LT = -1$ and the right panel is for $\LT = 1/2$. We again provide approximate boundaries for the region of parameter space where CR diffusion suppresses thermal instability (dashed lines). Note that these \textit{order-of-magnitude} boundaries are essentially the same as in the uniform medium case (see equations \ref{eq:weakdiff_cond} and \ref{eq:strongdiff_cond}, and Appendix \ref{app_diff} for a heuristic derivation). \eqref{eq:weakdiff_cond} and \eqref{eq:strongdiff_cond} are only slightly modified to emphasise the extra contribution coming from terms related to background CR heating, and are now:
\begin{equation}  \label{eq:weakdiff_cond_nonuni}
 | \frac{5}{2} - \LT | \frac{\omega_c}{\omega_d}  \lesssim  \eta \lesssim  | \LT |^{-1} \frac{\omega_d}{\omega_c} \ \ \ \quad \qquad (\omega_d < \omega_a)
\end{equation}
and
\begin{equation} \label{eq:strongdiff_cond_nonuni}
  | \frac{5}{2} - \LT | \frac{\omega_d \omega_c}{\omega_a^2}  \lesssim  \eta \lesssim | \LT |^{-1} \frac{\omega_d}{\omega_c} \ \ \qquad (\omega_d > \omega_a) 
\end{equation}
 respectively. If $\eta$ satisfies the above, thermal instability of modes with the corresponding $\omega_d$ and $\omega_a$ is suppressed by CR diffusion.  
 
As in Section \ref{sec:CR_field_length}, conditions \ref{eq:weakdiff_cond_nonuni} and \ref{eq:strongdiff_cond_nonuni} can be used to derive a CR Field length below which thermal instability is suppressed by CR diffusion. Like in the uniform medium, if $\kappa \omega_c / ( \eta v_A^2) \gtrsim 1$  then CR diffusion does not suppress thermal instability of high-$k$ modes and there is no associated CR Field length. If on the other hand $\kappa \omega_c / ( \eta v_A^2) \lesssim 1$, the CR Field length below which CR diffusion suppresses thermal instability is  approximately given by \eqref{eq:cr_field} (ignoring factors of order unity, e.g. $\propto \LT$). See Appendix \ref{app:crdiff_field} for more discussion.
 

\subsection{Thermal Stability versus Instability} \label{sec:nonuni_inst_cond}
We show $\LTC$ (the $\LT$ that is the boundary between overall thermal stability and instability, to any \textit{local} perturbation satisfying $\omega_a \gg \omega_c \eta^{-1} $) as a function of $\eta$ in the bottom panel of Figure \ref{fig:uni_boundary}. The solid black line again shows the boundary for $\kappa=0$, and the magenta dotted line is a broken power-law fit of the form
\begin{equation}
	\LTC = \frac{11}{6} [ 1+ \eta / \eta_*  ]^q,
\end{equation}
with $\eta_* = 1.19$ and $q = -1.13$ being the best-fit parameters. For a given $\eta$, the system is thermally unstable if $\LT < \LTC$. The dotted line shows the thermal stability boundary for perpendicular modes only, with $\omega_a = \omega_d = 0$ (note the lower plateau at small $\eta$, see eq. \ref{eq:etasmall_perp}).


The dashed horizontal line $\LTC = 5/2$ is the thermal stability/instability boundary if CR diffusion is present and $\kappa \gg \eta v_A^2 / \omega_c$. For $\kappa \gg \eta v_A^2 / \omega_c$ high-$k$ perturbations are isobaric and have a growth rate ${\rm Im}(\omega) = (2/5)\omega_c \Big( 5/2 - \LT  \Big)$ (Section \ref{sec:CR_field_length}). For $\kappa \lesssim \eta v_A^2 / \omega_c$, high-$k$ perturbations  are suppressed by CR diffusion. Only long-wavelength modes above the CR Field length can be thermally unstable. For $\kappa = v_A^2 / \omega_c$ (green dashed line) and $\eta > 1$, the CR Field length is at lower $k$ than the modes used for the stability boundary calculation ($\omega_a > 10 \omega_c \eta^{-1}, 10 \omega_c$) and so CR diffusion suppresses thermal instability of these modes. Perpendicular modes ($\bm{k \cdot B = 0}$, dotted line) are not affected by CR diffusion.

\section{CR Heating in gravitational field} \label{sec:gravity}
\subsection{Equilibrium}
As in Section \ref{sec:nonuni}, we consider equilibria in which cooling is completely balanced by cosmic-ray heating (equation \ref{eq:heat_bal}). Throughout this section, we neglect the effects of diffusion. Gravity, $\bm{g} = - g \bm{\hat{z}}$, changes the background gas pressure gradient to:
\begin{equation} \label{eq:dpg_gravity}
    \frac{dp_g}{d z} =\frac{d p_c } {d z}  \Big( \gamma \frac{\omega_{\rm ff}}{\omega_c} \frac{v_{A,z}}{c_s}-1 \Big),
\end{equation}
where $v_{A,z}$ is the z-component of the Alfv\'en velocity. We again assume a uniform magnetic field, $\bm{B} = B \sin \theta_B \ex$ + $B \cos \theta_B \ez$. 

\subsection{Thermal Instability}
 We find numerically (by solving equations \ref{eq:nonuni_drho}--\ref{eq:nonuni_pc} in MATLAB) that gravity does not significantly change thermal overstability growth rates for most modes (and it only slightly changes the entropy-mode oscillation frequency, which is dominated by the perturbed CR heating, i.e. $\omega_a$, unless $\omega_a < \omega_{\rm ff}$). In particular, for $\omega_a \gg \omega_{\rm ff}$ we recover the same growth rates as in Section \ref{sec:nonuni} and the growth rates obtained from equation \ref{eq:disp_1d} generally agree well with the exact calculation (which includes gravity). The green line in Figure \ref{fig:convection} shows this for $\LT = -1.0$, $\omega_{\rm ff}=20 \omega_c$ and $\omega_a \gg \omega_{\rm ff}$ ($\omega_a = 10^{3} \omega_c$): the growth rate is again $(2/5) \omega_c \Big( 11/6 - \LT \Big) $ at small $\eta$ and $- (2/3) \LT  \omega_c$ at large $\eta$.

 Gravity is more important when $\omega_a < \omega_{\rm ff}$ (e.g. modes with $\bm{k \cdot B} = 0$). This is shown by the blue and orange curves in Figure \ref{fig:convection} (with $\omega_a = 0$ and $\omega_{\rm ff} = 20 \omega_c$). At small $\eta$, gravity reduces the thermal instability growth rate by a factor of $\sim 2$ and the real part of the overstable entropy mode is dominated by the free-fall frequency, as has been found in previous work (\citealt{field65}).\footnote{For $\eta \ll 1$, gravity reduces the thermal instability growth rate to $(1/5) \omega_c \Big( 5/2 - \LT \Big)$, where the $5/2$ (instead of 2) arises from the perturbed CR heating term, $- \bm{\delta v_A \cdot \nabla} p_c \approx - (1/2) \omega_c p_g \delta \rho/\rho$.}
 
 \subsection{Convective Instability}
Figure \ref{fig:convection} also shows that there is a new form of instability occurring at larger $\eta$. We will show below that buoyancy is responsible for the increased growth rate. The buoyancy instability occurs only when  $\omega_a \lesssim \omega_{\rm ff}$, which corresponds to approximate adiabaticity. We note that convective behaviour in the presence of cosmic rays has been studied before by \cite{cd06}, \cite{dc09} and \cite{hz2018}. However, the setup we consider here, i.e. gravitationally stratified media with \textit{background CR heating}, was not part of their calculations. \cite{cd06} and \cite{dc09} did not include CR heating and focused on the effects of CR diffusion and thermal conduction (which tend to smooth out CR pressure and gas temperature along field lines, so their calculation differs substantially from our Schwarzschild-like calculation below). \cite{hz2018} looked at the effect of CR heating on the Parker instability. However, they did not consider \textit{background} CR heating, which is central to our buoyancy-instability calculation.  As a result, their instability calculation was different from the setup we consider here.

\subsubsection{Convective Instability Condition via Schwarzschild Criterion}
\begin{figure} 
  \centering
    \includegraphics[width = 0.45 \textwidth]{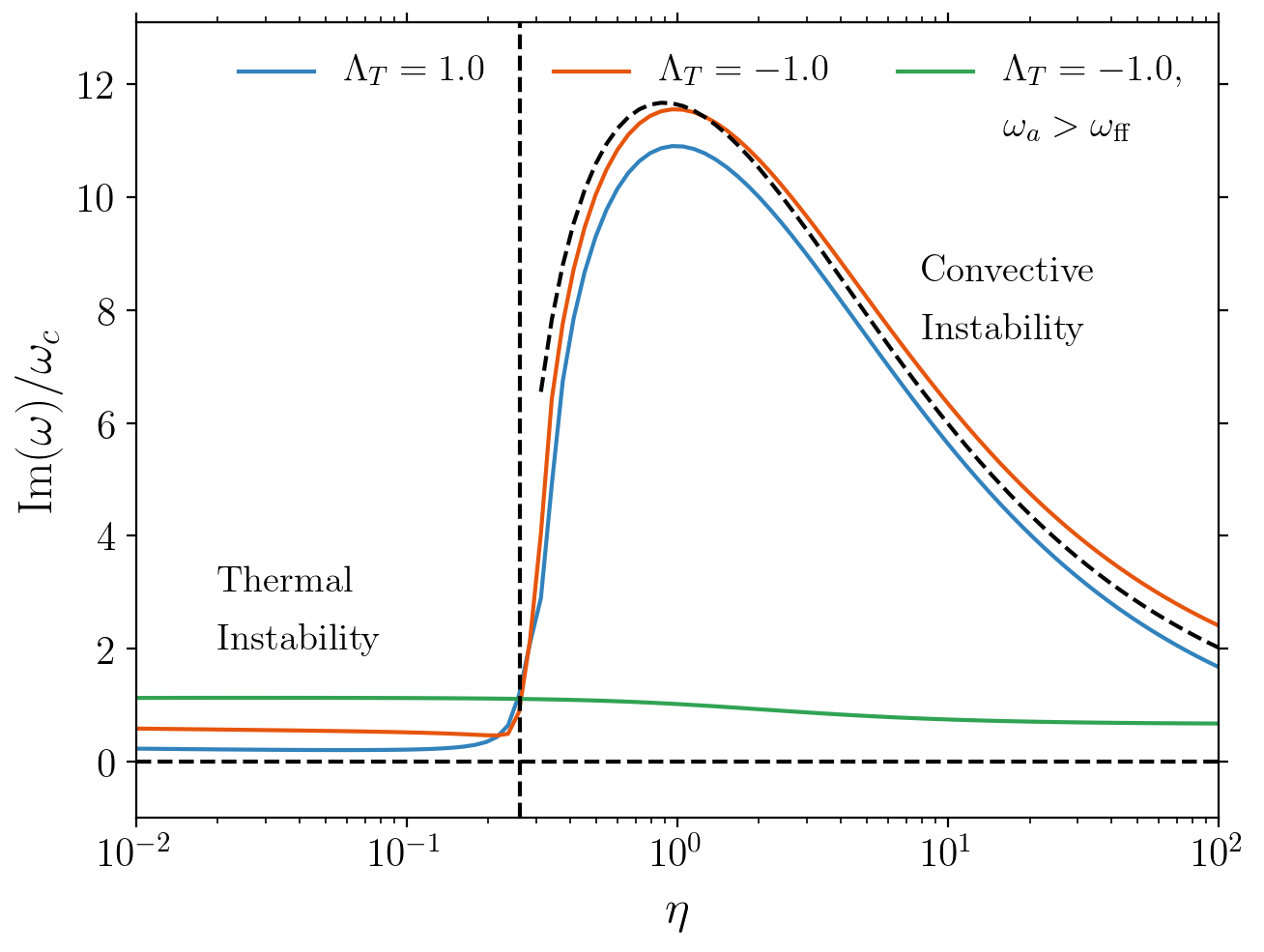}
  \caption{Thermal and convective instability of gravitationally stratified plasmas with $\omega_{\rm ff} = 20 \omega_c$, $\beta = 10$ and $\omega_a = 0$ (blue and orange lines). Buoyancy-driven instability occurs when $\eta$ satisfies eq. \ref{eq:schwarz_cond} (vertical dashed line). The dashed curve shows the approximate growth rate from equation \eqref{eq:conv_rate}. At smaller $\eta$ (and $\omega_a < \omega_{\rm ff}$), gravity reduces the thermal instability growth rate by a factor of $\sim 2$. When $\omega_a > \omega_{\rm ff}$ ($\omega_a = 10^3 \omega_c$; green line), perturbations are not  adiabatic and we recover the same thermal instability growth rate as without gravity.   \label{fig:convection}} 
\end{figure}

We can derive a convective stability criterion using the standard picture of a rising blob, which maintains pressure balance with its surroundings and is (approximately) adiabatic. For the latter, we require that $\omega_c < \omega_{\rm ff}$ (typically satisfied in galactic and cluster halos) and that $\omega_{a} < \omega_{\rm ff}$. The latter inequality is always satisfied for modes propagating perpendicular to the magnetic field, i.e. $\bm{k \cdot B}$ = 0. If both conditions are satisfied, then $\delta \ln ( p_c / \rho^{4/3}) \approx 0$ (from the CR pressure equation) and $\delta \ln ( p_g / \rho^{5/3}) \approx 0$ (from the gas entropy equation). In the high-$\beta$ limit, pressure balance and adiabaticity imply that:
\begin{equation}\label{eq:blob_pbal}
    \delta p_g + \delta p_c = \frac{5}{3} p_g \frac{\delta \rho}{\rho} + \frac{4}{3} p_c \frac{\delta \rho}{\rho}    = \bm{\xi \cdot \nabla } p_g + \bm{\xi \cdot \nabla } p_c .
\end{equation}
The displaced fluid element will be buoyantly unstable if $\delta \rho  < \bm{\xi \cdot \nabla} \rho$, so the condition for instability is
\begin{equation}
    \bm{\xi \cdot \nabla } p_g + \bm{\xi \cdot \nabla } p_c <   \frac{5}{3} \frac{p_g}{\rho}  \bm{\xi \cdot \nabla } \rho +\frac{4}{3} \frac{p_c}{\rho} \bm{\xi \cdot \nabla } \rho.
\end{equation}
Using \eqref{eq:dpg_gravity} and that the background density and CR pressure satisfy $\rho \propto p_c^{3/2}$, this can be rewritten as:
\begin{equation}  \label{eq:schwarz_cond}
    \eta \Big(\gamma \frac{ \omega_{\rm ff}}{\omega_c}\frac{v_{A,z}}{c_s} -2 \Big) > \frac{3}{2} \gamma,
\end{equation}
where $\gamma= 5/3$ is the gas adiabatic index. 
We derive the same criterion directly from the linearised equations in Appendix \ref{app:conv} (also assuming adiabaticity). The above condition turns out to be equivalent to
\begin{equation} \label{eq:eff_s}
    \frac{d s_{\rm eff}}{dz} \propto \frac{d }{dz} \Big( \ln \frac{p_g}{\rho^{5/3}} + \eta \ln \frac{p_c}{\rho^{4/3}} \Big) < 0.
\end{equation}
If the above is satisfied, the system is convectively unstable. Condition \eqref{eq:schwarz_cond} is shown in Figure \ref{fig:convection} as the dashed vertical line.

\begin{figure} 
  \centering
    \includegraphics[width = 0.45 \textwidth]{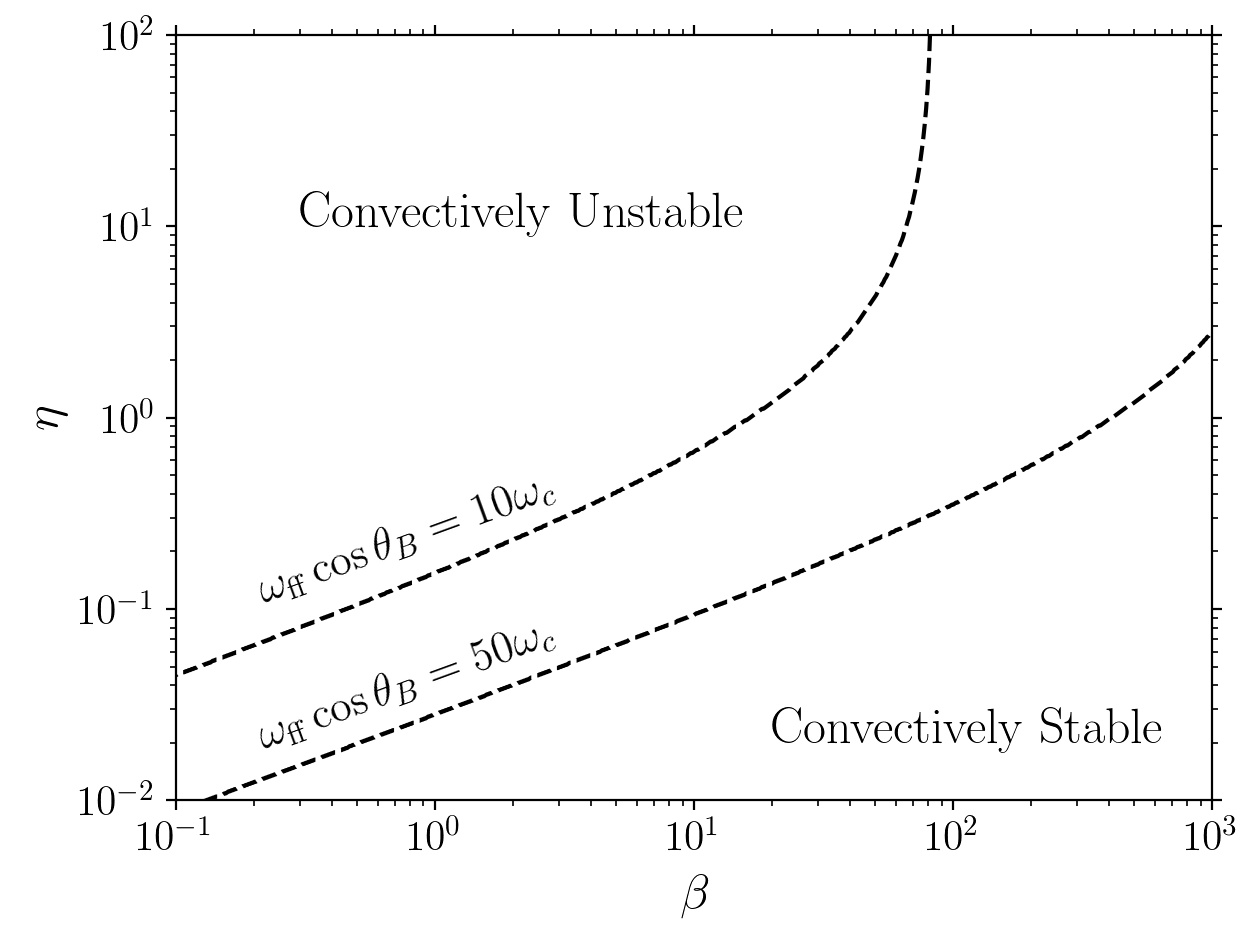}
  \caption{ Convective stability/instability boundary in the ($\beta$, $\eta$) plane for two choices of $\omega_{\rm ff} \cos \theta_B / \omega_c$, where $\theta_B$ is the angle between the z-axis (direction of gravity and pressure gradients) and the magnetic field. For a given choice of $\omega_{\rm ff} \cos \theta_B / \omega_c$, there is a maximum $\beta$ for which convection can occur, as determined by equation \eqref{eq:schwarz_cond}.  \label{fig:convective_map}} 
\end{figure}

Using hydrostatic equilibrium (\ref{eq:dpg_gravity}), we can further rephrase the instability criterion in terms of the CR and gas pressure scale heights ($H_c^{-1} \equiv d \ln p_c / d z$, $H_g^{-1} \equiv d \ln p_g / d z$),
\begin{equation} \label{eq:schwarz_Hc}
 \frac{H_c}{H_g} - \eta > \frac{3}{2} \gamma .
\end{equation}
Therefore, a necessary condition for convection is that $H_c / H_g > 5/2$. We show the convective (in)stability in the ($\eta$, $\beta$) plane in Figure \ref{fig:convective_map}. The system becomes convectively unstable for a larger range of $\eta$ and $\beta$ when  $\omega_{\rm ff} \cos \theta_B / \omega_c$  is increased.

We derive an approximate growth rate for the convective instability in the limit $\omega_{\rm ff} \gg \omega_c, \omega_a$ in Appendix \ref{app:conv}. The approximate growth rate (equation \ref{eq:conv_rate}, which is derived by dropping any dependence on $\LT$) is shown in Figure \ref{fig:convection} as the dashed line and agrees well with the exact solution. 

\section{Discussion}\label{sec:conclusions}
In this paper, we have studied the linear thermal stability of systems heated by streaming cosmic rays. Streaming cosmic rays can be an important heating mechanism in cluster halos (\citealt{guo08}; \citealt{jp_1}; \citealt{jp_2}). Our order-of-magnitude estimates suggest that CR heating may also be important in galactic halos, particularly for Milky Way mass systems (see Section \ref{sec:balcool} and Figure \ref{fig:heat_vs_cool}). Thermal instability is a viable mechanism for creating the cold gas that is found in these systems. 

We calculated thermal instability growth rates in the presence of CR heating for a wide range of $\eta \equiv p_c /p_g$, in three background equilibria: in a uniform background (where cooling is balanced by an unspecified volumetric  heating, but CR heating is present in the perturbed equations; Section \ref{sec:uniform}), in a background with CR heating balancing cooling (Section \ref{sec:nonuni}), and in a gravitationally stratified background heated by CRs (Section \ref{sec:gravity}). Gas in PIE in galactic halos is a special case of our uniform background calculation (Section \ref{sec:pie}), which is always thermally stable, independent of CR pressure.

The key timescales/frequencies in this problem are summarised in equations \ref{eq:wc}--\ref{eq:wff} and include the cooling, sound, Alfv\'en (CR heating), CR diffusion and free-fall frequencies. The Alfv\'en frequency enters the thermal instability calculation primarily via the perturbed CR heating, as the entropy mode is not sensitive to magnetic tension and pressure. We focused on local WKB perturbations, satisfying equations \ref{eq:w_s_vs_cff} and \ref{eq:w_a_vs_cff}, in the high-$\beta$ limit. Our fiducial parameter set was $\omega_a = 10^{3}\omega_c$, $\beta = 100$, and $\omega_{\rm ff} = 20 \omega_c$ when we included gravity (this is well motivated in galactic and cluster halos, where typically $\omega_{\rm ff} \gtrsim 10 \omega_c$, but we stress that it need not be true in general astrophysical systems). Our results depend weakly on $\beta$ for $\beta \gtrsim 3$. Thermal instability growth rates also do not depend strongly on wavenumber $k$ for $\omega_a \gtrsim \omega_c$  ($\omega_a \gtrsim \omega_c \eta^{-1}$)  in the uniform  (CR-heated) background, which corresponds to WKB perturbations satisfying $kH \gtrsim 1$. Figure \ref{fig:uni} shows this weak dependence for $\omega_a \leq 10^{3} \omega_c$ (left panels) and $3\leq \beta \leq 100$ (middle panels).           

We have focused on cosmic rays that stream down their pressure gradient at the Alfv\'en velocity, while also including CR diffusion along the magnetic field. We find that CR diffusion can suppress thermal instability of a subset of modes (right panels of Figure \ref{fig:uni} and Figure \ref{fig:uni_map}) and modify the overall thermal stability to arbitrary perturbations (Figure \ref{fig:uni_boundary}). However, the dominant CR transport process in galactic halos remains unclear and it is uncertain whether CR streaming and diffusion are generally simultaneously relevant. Indeed, a variety of work suggests that if CR scattering is mostly due to self-excited Alfv\'en waves, then even cosmic rays that are imperfectly coupled to the thermal plasma (where the imperfect coupling is what gives rise to a possibly diffusive behaviour) are not truly diffusive (\citealt{skilling71}; \citealt{wiener2013}; \citealt{wiener18}), and instead stream at super-Alv\'enic speeds. Super-Alfv\'enic streaming does not, however, increase the rate at which CRs heat the gas (i.e. the CRs still heat the gas at a rate $- \bm{v_A \cdot \nabla}p_c$). Moreover, how super-Alfv\'enic streaming speeds depend on other fluid quantities is not well known. As a result, we did not include super-Alfv\'enic streaming in our calculation.

The different background equilibria we have considered allowed us to disentangle how CR physics affects thermal instability. Independent of background, thermal instability growth rates depend strongly on $\eta$, which determines whether the instability is isobaric (small $\eta$) or isochoric (large $\eta$); see equation \ref{eq:perts} and associated discussion. The perturbed CR heating also introduces high-frequency  oscillations (order $\omega_a$ for $\eta \sim 1$, independent of background) in the gas entropy mode (see Section \ref{sec:uni_streaming}), so that thermal instability is formally an overstability with an oscillation frequency comparable to or larger than its growth rate. This is true even in a uniform medium, where thermal instability is normally a purely growing mode. CR heating does not, however, modify isobaric (small $\eta$) or isochoric (large $\eta$) thermal instability growth rates, $(2/5) \omega_c (2-\LT)$ and $-(2/3)\LT \omega_c$ respectively, in a uniform background (Section \ref{sec:uni_asymptotes} and top panels of Figure \ref{fig:uni}). Background CR heating does slightly change isobaric thermal instability growth rates at small $\eta$ (Sections \ref{sec:nonuni_1d} and \ref{sec:perp_modes}, and bottom panels of Figure \ref{fig:uni}). Incorporating gravity in our analysis (Section \ref{sec:gravity}) did not significantly affect thermal instability growth rates, but it allowed us to determine under what conditions a gravitationally stratified, CR-heated medium is buoyantly unstable.  

 Thermal instability growth rates as a function of $\eta$ are plotted in Figure \ref{fig:uni} for different cooling curve slopes $\LT$ (left panels), plasma-$\beta$ (middle panels) and for different CR-diffusion frequencies $\omega_d$ (right panels; there is no CR diffusion present in the left and middle panels). The top panels show the uniform-background (Section \ref{sec:uniform}) calculation results, the bottom panels show the corresponding results in a background in which cosmic-ray heating balances cooling (Section \ref{sec:nonuni}). As already mentioned in the previous paragraph, thermal instability growth rates depend strongly on $\eta$. Growth rates do not depend strongly on $\beta$ for $\beta \gtrsim 3$, as is expected in galaxy halos, groups and clusters.

Figure \ref{fig:uni_boundary} shows the boundary between thermal stability and instability to arbitrary (WKB and high-$\beta$) perturbations in a uniform background (top) and the CR-heated background (bottom). We formulate this in terms of the critical cooling curve slope  $\LTC = \partial \ln \Lambda / \partial \ln T$ above which all perturbations are thermally stable. When there is no CR diffusion present, the ``critical" cooling curve slope $\LTC$ as a function of $\eta$ is well described by a broken power law. If CR diffusion is present and $\kappa \omega_c / \eta v_A^2 \gg 1$, the stability/instability boundary is simple and independent of $\eta$: CR diffusion renders high-$k$ (large $\omega_d$, $\bm{k\cdot B}\neq 0$) perturbations isobaric, so that $\LTC = 2$ in a uniform background (eq. \ref{eq:small_eta_uni}) and $\LTC = 5/2$ in the CR-heated background (eq. \ref{eq:small_eta_nonuni}). If $\kappa \omega_c / \eta v_A^2 \lesssim 1$, CR diffusion introduces a Field length below which thermal instability is suppressed (eq. \ref{eq:cr_field}). This affects the thermal stability/instability boundary in Figure \ref{fig:uni_boundary} (e.g., green dashed line for $\kappa= v_A^2 / \omega_c$). We stress, however, that unlike thermal conduction, there is only an effective CR Field length for particular CR diffusion coefficients, namely $\kappa \lesssim \eta v_A^2 / \omega_c$. For $\kappa \rightarrow \infty$ CRs have no effect on thermal instability because the CR pressure is essentially uniform.

Modes with $\omega_{\rm ff} > \omega_{\rm a}$ (i.e. nearly adiabatic modes) can further be convectively unstable (driven by buoyancy). Convective instability occurs if equation \ref{eq:schwarz_cond} is satisfied (see also Figure \ref{fig:convective_map}). By defining $s_{\rm eff} \propto \ln p_g / \rho^{5/3} + \eta \ln p_c / \rho^{4/3}$, the criterion for convective instability can be written in the form $d s_{\rm eff}/dz <0 $. In our setup, this turns out to be satisfied if the ratio of the CR pressure scale height to the gas pressure scale height is sufficiently large (equation \ref{eq:schwarz_Hc}). We also derive an approximate expression for the growth rate of the convective instability for perpendicular modes (see equation \ref{eq:conv_rate} and the dashed curve in Figure \ref{fig:convection}). Our calculation differs from previous work that considered buoyancy instabilities in the presence of cosmic rays, which did not consider \textit{background} CR heating (\citealt{cd06}; \citealt{dc09}; \citealt{hz2018}). Background CR heating is essential in our calculation, as it is the background gas-pressure gradient, set by hydrostatic equilibrium and $\bm{\nabla}p_c$ (which is set by cooling), that drives convection.


Our calculations show that systems heated by cosmic rays are likely thermally unstable for temperature ranges relevant to galactic halos ($10^{5} {\rm K} \lesssim T \lesssim 10^{7} {\rm K}$, where $\LT \lesssim 0$; \citealt{draine}). In halos that are in PIE, however, $\LT$ is large ($>2$) and the gas is thermally stable for any $\eta$. In cluster halos, where the temperature can exceed $\approx 10^{7}K$ and thermal Bremsstrahlung is the dominant radiative cooling process (with $\Lambda_T = 0.5$), CR heating could lead to thermal stability if CR pressure dominates (i.e. $\eta \gtrsim 1$) and CR streaming dominates over diffusion. If CR diffusion is important (and $\kappa \gg \eta v_A^2 / \omega_c$), however, it eliminates CR pressure perturbations and cooling by Bremsstrahlung would be thermally unstable. Moreover, $\eta \gtrsim 1$ in cluster halos is disfavoured observationally (e.g., \citealt{huber2013}). It is thus likely that all halo gas in CIE is thermally unstable in the presence of CR heating.
 
It remains to be seen how CR heating affects the nonlinear evolution of the thermal instability and the resulting multiphase structure of halo gas. In particular, are there significant differences introduced by the $\mathcal{O}(\omega_a)$ entropy oscillations introduced by the CR heating term? This heating frequency can be larger than the free-fall frequency, and it is plausible that this may change the effect of buoyant oscillations on the saturation of thermal instability. However, we note the caveat that long-wavelength modes tend to dominate the nonlinear saturation of thermal instability, for which $\omega_a > \omega_{\rm ff}$ is not necessarily satisfied. A sufficiently small ratio of the cooling time to the free-fall time, $t_{\rm cool} / t_{\rm ff} \lesssim 10$, has been identified as crucial for the development of multiphase gas in hydro simulations (e.g. \citealt{smqp12}). \cite{ji18} showed that magnetic fields enhance thermal instability by suppressing buoyant oscillations via magnetic tension. Future simulations will address how entropy oscillations driven by CR heating (which also occur at $\sim$ the Alfv\'en frequency) affect this evolution and the creation of multiphase gas. In particular, it seems plausible that the dimensionless ratios $t_{\rm cool} / t_{\rm A}$ (with $t_{\rm A} \equiv H / v_A$) and $\eta = p_{c}/p_g$, which are related to the propagation speed of thermally unstable modes, may be important for the nonlinear evolution of thermal instability.

\section*{Acknowledgements}
We thank S. P. Oh, Y. Jiang, X. Bai, P. Hopkins, S. Ji, M. Kunz, E. Ostriker, C. Pfrommer, A. Spitkovsky, J. Squire \& E. Zweibel for enlightening discussions. This research was supported in part by the Heising-Simons Foundation, the Simons Foundation, and National Science Foundation Grant No. NSF PHY-1748958, NSF grant AST-1715070,  and a Simons Investigator award from the Simons Foundation. EQ thanks the Princeton Astrophysical Sciences department and the theoretical astrophysics group and Moore Distinguished Scholar program at Caltech for their hospitality and support. PK would like to thank the Kavli Institute for Theoretical Physics for their hospitality and support offered via the Graduate Fellowship Program.

\bibliographystyle{mnras}
\bibliography{crti}


\appendix

\section{Cosmic-Ray Pressure Fraction} \label{app:eta}
In this section we derive order-of-magnitude estimates of the CR pressure fraction, $\eta$, which we used to create the bottom panel of Figure \ref{fig:heat_vs_cool}. We consider the injection of cosmic rays by Type II Supernovae (Section \ref{sec:typeII}) and AGNs (Section \ref{sec:agn}). This enables us to estimate the (global) cosmic ray energy budget, which is related to a spatially-averaged CR pressure fraction $\eta$. We find that both mechanisms can in principle produce a significant cosmic-ray pressure, with the caveat that we treat the system in a globally-averaged sense.

\subsection{Gas Thermal Energy}
We want to compare the total cosmic ray energy to the total thermal energy of the gas within the halo. In what follows, we estimate the thermal energy of the gas filling the galaxy out to the virial radius $R_{\rm vir}$. We define $R_{\rm vir}$ as the radius within which the mean matter density is 200 times the cosmic critical density, i.e. $\langle \rho_m \rangle _{R_{\rm vir}} = 200 \rho_c$.  We can approximate the total thermal energy within the virial radius as
\begin{equation}
    E_{\rm th}  \sim \frac{3}{2} \int_0^{R_{\rm vir}} \frac{ \rho_g k_B T }{ m_H}  4 \pi r^2 dr
\end{equation}
where $\rho_g$ is the thermal gas density. We assume an isothermal profile with virialized $k_B T = GM_{200} m_H/ (3R_{\rm vir})$:
\begin{equation}
     E_{\rm th}  \sim 2 \pi R_{\rm vir}^3 \frac{\langle \rho_g \rangle_{R_{\rm vir}} k_B T }{m_H}  \sim  \frac{\langle \rho_g \rangle _{R_{\rm vir}}}{ \langle \rho_m\rangle _{R_{\rm vir}} } \frac{GM_{200}^2}{2R_{\rm vir}},
\end{equation}
where $M_{200} = (4 \pi /3) \langle \rho_m \rangle R_{vir}^3$. For $\langle \rho_g \rangle _{R_{\rm vir}} = (x \Omega_B / \Omega_M) \langle \rho_m \rangle_{\rm vir} $, where $x$ accounts for missing baryonic mass relative to the cosmic mean, we obtain
\begin{equation} \label{eq:Eth}
    E_{\rm th} \sim  \frac{x \Omega_B}{\Omega_M} \frac{G M_{200}^2}{2 R_{\rm vir}} \sim 3.6 \times 10^{58} \  \mathrm{ergs}  \ \Big( \frac{x}{1.0} \Big) \Big( \frac{M_{200}}{10^{12} M_{\odot}} \Big)^{5/3} .
\end{equation}

\subsection{Energy of Cosmic Rays: Injection by Type II SNe \label{sec:typeII}} 
We first consider cosmic rays injected by Type II Supernovae. 
We assume that there is a core-collapse Supernova for every $100 M_{\odot}$ of stars formed, so that $N_{\rm II}=M_{*}/100M_{\odot}$ is the total number of Type II SNe in a galaxy with stellar mass $M_{*}$. We define $\chi$ as the ratio of the number of cosmic rays still present in the halo (out to the virial radius) to the total number produced throughout the galaxy's lifetime. The total cosmic ray energy in the halo is then
\begin{equation} \label{eq:ecr_sym}
E_{c} = \chi f_{\rm II} E_{\rm II} N_{\rm II}.
\end{equation}
In the above expression, $f_{\rm II} E_{\rm II}$ is the typical CR energy injected by a single type II Supernova ($f_{\rm II}$ is the fraction of supernova energy released as cosmic rays). As is commonly assumed, we take $f_{\rm II} \approx 0.1$ (\citealt{zweibel17}), so that for a typical non-neutrino energy of $10^{51}$ ergs released by Type II Supernovae, $f_{\rm II} E_{\rm II} \approx 10^{50} {\rm ergs}$. This implies that
\begin{equation} \label{eq:Ecr_sn}
E_{c}  \sim  10^{59}  \  \mathrm{ergs} \ \Big( \frac{\chi}{1.0} \Big) \Big( \frac{M_{*}}{10^{11} M_{\odot}} \Big),
\end{equation}
where our choice of $\chi = 1.0$ reflects the possibility that in massive galaxies with large $R_{\rm vir}$,  the escape time of cosmic rays may be of order the galaxy age. Comparing \eqref{eq:Ecr_sn} to \eqref{eq:Eth} gives
\begin{equation}
\eta \sim \frac{E_{c}}{2 E_{\rm th}} \sim 1.4 \ \Big( \frac{\chi}{1.0} \Big) \  \Big( \frac{x}{1.0} \Big)^{-1}  \Big(  \frac{M_{*}}{10^{11} M_{\odot}} \Big)  \Big( \frac{M_{200}}{10^{12} M_{\odot}} \Big)^{-5/3}    .
\end{equation}

\subsection{Energy of Cosmic Rays: Injection by SMBHs \label{sec:agn}} 
We now consider cosmic rays that are created by SMBHs. We can estimate the CR energy content by assuming that a fraction $f_{\rm BH}$ of the SMBH luminosity goes into cosmic rays, i.e. the SMBH injects cosmic-ray energy at a rate $ \epsilon f_{\rm BH} \dot{M} c^2$, where $\epsilon$ is the black hole's radiative efficiency and $\dot{M}$ is its mass accretion rate. This gives:
\begin{equation} 
E_{c} \sim \chi \epsilon f_{\rm BH}  M_{\rm BH} c^2,
\end{equation}
where $\chi$ is again defined as the ratio of the number of cosmic rays still present in the halo to the total number produced throughout the galaxy's lifetime. The total CR energy is approximately
 \begin{equation}
    E_{c} \sim  18  \times 10^{59} {\rm ergs} \  \Big( \frac{\chi}{1.0} \Big) \ \Big( \frac{ \epsilon f_{\rm BH}}{10^{-3}} \Big) \Big( \frac{M_{\rm BH}}{10^9 M_{\odot}} \Big)  .
\end{equation}
Comparing this to the total thermal energy in equation \eqref{eq:Eth}, we find that
\begin{equation}
\eta \sim \frac{E_{c}}{2 E_{th}} \sim 25  \ \Big( \frac{\chi}{1.0} \Big) \ \Big( \frac{ \epsilon f_{\rm BH}}{10^{-3}} \Big) \ \  \Big( \frac{x}{1.0} \Big)^{-1}  \Big( \frac{M_{\rm BH}}{10^9 M_{\odot}} \Big)  \Big( \frac{M_{200}}{10^{12} M_{\odot}} \Big)^{-5/3}  .
\end{equation}



\section{Cosmic-ray diffusion versus thermal instability}
\label{app_diff}
We now provide a short, heuristic derivation for conditions \eqref{eq:weakdiff_cond} and \eqref{eq:strongdiff_cond}. For simplicity, here we consider the case of a uniform background (Section \ref{sec:uniform}). An analogous calculation for the CR-heated background gives the very similar conditions \eqref{eq:weakdiff_cond_nonuni} and \eqref{eq:strongdiff_cond_nonuni} (see also Figure \ref{fig:uni_map}). In \ref{app:crdiff_field} we show that CR diffusion can introduce a Field length, below which thermal instability is suppressed.

 \subsection{Modes with $\omega_d \ll \omega_c$}
 In the limit where $\omega_d = 0$ or $\omega_d \ll \omega_c$, diffusion is negligible for any $\eta$ and the $\eta \rightarrow 0$, $\eta \rightarrow \infty$ limits are connected smoothly at intermediate $\eta$ (as in Figure \ref{fig:uni}).  \\
 
  \subsection{Modes with $\omega_c \ll \omega_d \lesssim \omega_a $ } \label{sec:app_weakdif}
In this limit, \eqref{eq:uni_pc} gives an approximate leading-order relation between $\delta p_c$ and $\delta \rho$:
 \begin{equation} \label{eq:pc_vs_rho_weakdiff}
     \frac{\delta p_c}{p_g} \sim  \eta \frac{\delta \rho}{\rho} \Big( 1 + i \frac{\omega_d}{\omega_a} \Big).
 \end{equation}
 Inserting this approximate relation into equation \ref{eq:uni_pg} gives
 \begin{equation} \label{eq:uni_pg_app}
    \frac{\delta p_g}{p_g} \Big( \frac{\omega}{\gamma - 1} + i \omega_c \LT   \Big) - \frac{\delta \rho}{\rho} \Big( \frac{\gamma \omega}{\gamma-1} - i \omega_c  (2 - \LT ) \Big)  \sim  \frac{\delta \rho}{\rho} \eta (\omega_a + i \omega_d).
\end{equation}
For $\eta \ll 1$, i.e. $\delta p_g / p_g \approx - \delta p_c / p_g \ll \delta \rho / \rho$, the CR-diffusion term introduced by the perturbed CR heating essentially acts like a thermal-conduction term with a thermal diffusion coefficient $\sim \eta \kappa$. The diffusive term $\propto i \eta \omega_d \delta \rho / \rho$ acts to oppose the perturbed cooling term $\propto i \omega_c (2 - \LT) \delta \rho / \rho$ which drives thermal instability. Diffusion suppresses thermal instability if:
 \begin{equation} \label{eq:weakdiff_lowbound}
     \eta \omega_d  \gtrsim |2 - \LT | \ \omega_c ,
 \end{equation}
which gives the lower bound in \eqref{eq:weakdiff_cond}. 

This suppression of thermal instability by CR diffusion is not present at large $\eta$, when the second term on the LHS of \eqref{eq:uni_pg_app}, $\propto \delta \rho / \rho$, is negligible (thermal instability is isochoric). Using $\delta p_g \approx - \delta p_c$ and  \eqref{eq:pc_vs_rho_weakdiff} in eq. \ref{eq:uni_pg_app} one can show that CR diffusion is unimportant when
\begin{equation} \label{eq:weakdiff_upperbound}
   \eta \gtrsim \frac{\omega_d}{\omega_c} |\LT|^{-1},
\end{equation}
 at which point we recover the $\eta \rightarrow \infty$ (isochoric) asymptotic growth rate (eq. \ref{eq:etainfy}).
 
  \subsection{Modes with $\omega_d \gg \omega_a $ }
In this limit, \eqref{eq:uni_pc} gives:
 \begin{equation} \label{eq:app_diff_1}
     \frac{\delta p_c}{p_g} \sim  \frac{\delta \rho}{\rho} i \eta \frac{\omega_a}{\omega_d}.
 \end{equation}
 Inserting this into equation \ref{eq:uni_pg} gives
 \begin{equation} \label{eq:uni_pg_app2}
    \frac{\delta p_g}{p_g} \Big( \frac{\omega}{\gamma - 1} + i \omega_c \LT   \Big) - \frac{\delta \rho}{\rho} \Big( \frac{\gamma \omega}{\gamma-1} - i \omega_c  (2 - \LT ) \Big)  \sim  \frac{\delta \rho}{\rho} i \eta \frac{\omega_a^2}{\omega_d} .
\end{equation}
Note that the CR heating term $\propto \eta \omega_a^2 / \omega_d = \eta v_A^2 / \kappa$ is scale independent. CR diffusion again acts to oppose the perturbed cooling term $\propto i \omega_c (2 - \LT) \delta \rho / \rho$ which drives thermal instability. For $\delta p_g / p_g \ll \delta \rho / \rho$ (for $\eta \ll \omega_d / \omega_a$) CR diffusion 
suppresses thermal instability if
\begin{equation}\label{eq:strongdiff_lowbound}
    \eta \frac{\omega_a^2}{\omega_d} \sim  \frac{\eta v_A^2}{\kappa} \gtrsim | 2 - \LT |\omega_c.
\end{equation} 
This is the lower bound in \eqref{eq:strongdiff_cond}. Note that there is no scale dependence. As a result, if $\kappa \omega_c / (\eta v_A^2) \gg 1$ then short-wavelength modes with $\omega_d \gg \omega_a$ are not suppressed by CR diffusion. If, however, $\kappa \omega_c / (\eta v_A^2) \ll 1$, CR diffusion instead leads to the decay of high-$k$ gas-entropy modes (see Figure \ref{fig:diff_vs_k}). 

As described before in \ref{sec:app_weakdif}, when $\eta$ is sufficiently large for the second term on the LHS to be negligible (thermal instability is isochoric), CR diffusion does not affect the TI growth rate. CR diffusion is unimportant when:
\begin{equation} \label{eq:strongdiff_upperbound}
   \eta \gtrsim \frac{\omega_d}{\omega_c} |\LT|^{-1}.
\end{equation}
This is the upper bound in \eqref{eq:strongdiff_cond}.
 
\subsection{CR-Diffusion Field Length} \label{app:crdiff_field}
 \begin{figure} 
  \centering
    \includegraphics[width = 0.45\textwidth]{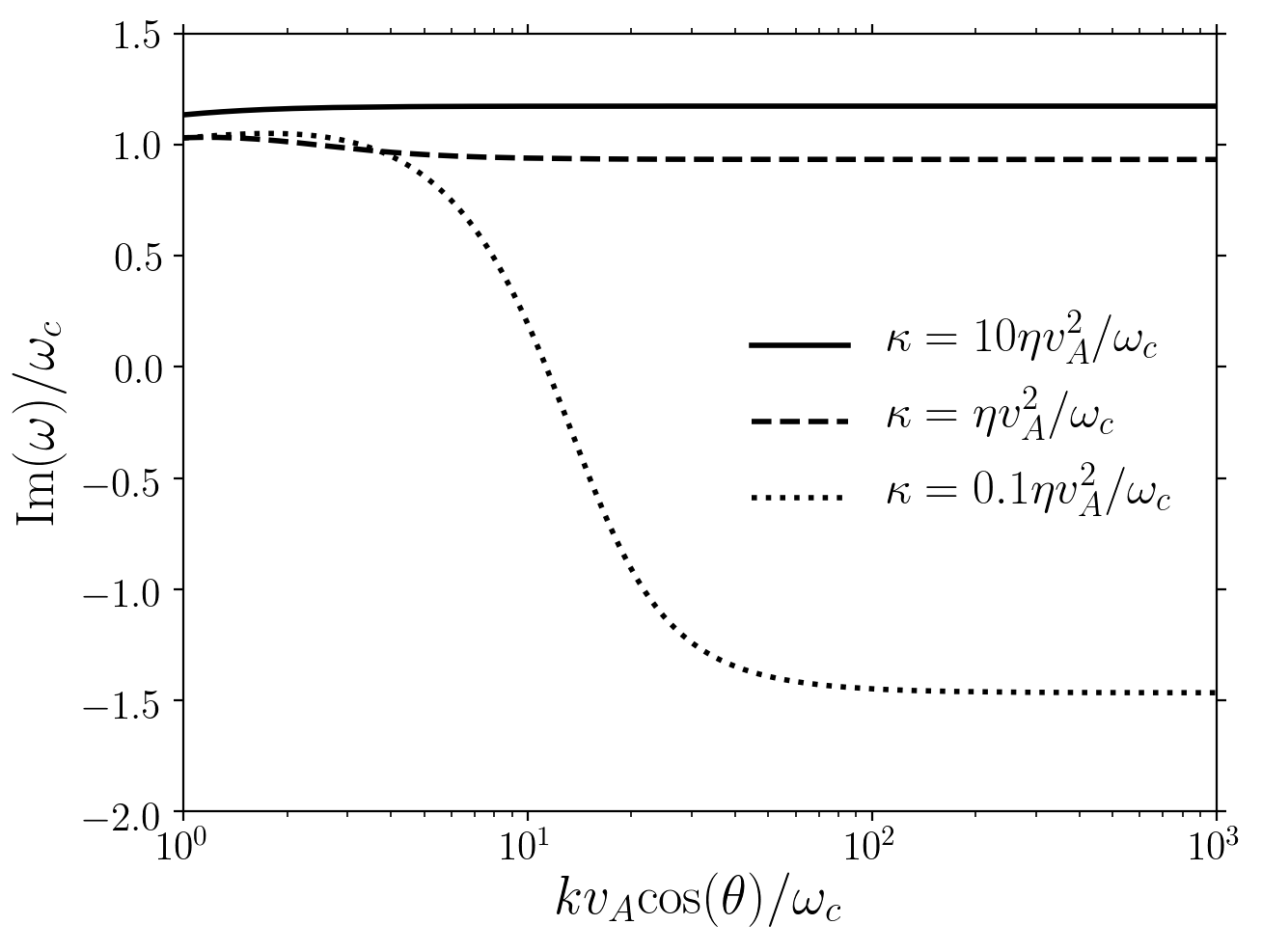}
  \caption{Thermal instability growth rates as a function of wavenumber $k$ for $\LT=-1$, $\beta = 100$ and $\eta = 1$ in a uniform medium. $\theta$ is the angle between $\bm{k}$ and the background magnetic field. We show growth rates for different CR diffusion coefficients $\kappa$. For $\kappa \gg \eta v_A^2 / \omega_c$ (solid line) CR diffusion does not affect thermal instability growth rates at high $k$. For $\kappa \ll \eta v_A^2 / \omega_c$ (dotted line) diffusion introduces a CR Field length below which thermal instability is suppressed. In all cases, growth/damping rates are constant at high $k$, as the perturbed CR heating is scale-independent for $\omega_d \gg \omega_a$.  \label{fig:diff_vs_k}} 
\end{figure}
 We can rephrase conditions \ref{eq:weakdiff_cond} (eq. \ref{eq:weakdiff_lowbound} and eq. \ref{eq:weakdiff_upperbound}) and \ref{eq:strongdiff_cond} (eq. \ref{eq:strongdiff_lowbound} and eq. \ref{eq:strongdiff_upperbound}) in terms of length scales at which CR diffusion suppresses thermal instability. In particular, in Appendix \ref{sec:app_weakdif} we demonstrate that CR diffusion can play a role similar to thermal conduction. This suggests that there is a CR-diffusion analogue of the Field length for thermal conduction (\citealt{field65}) below which thermal instability is suppressed. Thermal instability of long-wavelength modes with $\bm{\hat{b} \cdot k} < v_A / \kappa$ ($\omega_d < \omega_a$) is suppressed by CR diffusion if
\begin{equation} \label{eq:uni_weakdiff_2}
    (\bm{\hat{b} \cdot k})^2 \gtrsim {\rm max} \Big( \frac{ \omega_c}{\eta \kappa} |2 - \LT|, \frac{\eta \omega_c}{\kappa} |\LT| \Big).
\end{equation}
The above is derived from and equivalent to equation \ref{eq:weakdiff_cond}.
Thermal instability of short-wavelength modes with rapid CR diffusion, $\bm{\hat{b} \cdot k} > v_A / \kappa$ ($\omega_d > \omega_a$), is suppressed by CR diffusion if
\begin{equation} \label{eq:uni_strongdiff_2}
    \frac{\kappa \omega_c}{\eta v_A^2}|2-\LT| \lesssim 1 \ \ {\rm and} \ \ (\bm{\hat{b} \cdot k})^2 \gtrsim \frac{\eta \omega_c}{\kappa} |\LT|.
\end{equation}
The above is equivalent to equation \ref{eq:strongdiff_cond}. 

Note that if $\kappa \gtrsim \eta v_A^2 / \omega_c$ ($\omega_d \gtrsim \eta \omega_a^2 / \omega_c$) then CR diffusion does not suppress thermal instability of high-$k$ modes ($\omega_d \gg \omega_a$), even though $\omega_d$ is large. There is therefore no ``CR Field length" below which thermal instability is completely suppressed. Instead, the instability of high-$k$ modes is isobaric with growth rates ${\rm Im}(\omega) = (2/5) \Big(2 - \LT\Big) \omega_c$. Cosmic rays have no effect on thermal instability as the rate at which they heat the gas at high $k$ is less than the cooling rate ($\kappa \omega_c / \eta v_A^2$ is the ratio of the cooling rate to the CR heating rate at high $k$, see eq. \ref{eq:uni_pg_app2}). 

Conversely, if $\kappa \lesssim \eta v_A^2 / \omega_c$, the cosmic-ray heating rate at high $k$ exceeds the gas cooling rate. CR diffusion then suppresses thermal instability of high-$k$ gas-entropy modes. In other words, when $\kappa \lesssim \eta v_A^2 / \omega_c$ there is a maximum $\bm{\hat{b} \cdot k}$ at which thermal instability occurs. Using \eqref{eq:uni_weakdiff_2} and \eqref{eq:uni_strongdiff_2} and dropping order unity $\sim \LT$ factors one can show that the CR Field length is given by:
\begin{equation} \label{eq:cr_field_app}
  \lambda_{\rm CRF} \sim
    \begin{cases}
      2 \pi |\bm{\hat{b} \cdot \hat{k}}|  \sqrt{\frac{\eta \kappa}{\omega_c}} & \eta < 1\\
            2 \pi |\bm{\hat{b} \cdot \hat{k}}|  \sqrt{\frac{ \kappa}{\eta \omega_c}} & \eta > 1.
    \end{cases}       
\end{equation}
This is the CR-diffusion analogue of the Field length (\citealt{field65}). We stress again that this CR Field length exists only if $\kappa \lesssim \eta v_A^2 / \omega_c$. Figure \ref{fig:diff_vs_k} shows how the value of $\kappa $ determines the stability/instability of high-$k$ modes.

\section{Linearised Equations with Background Cosmic-Ray Heating }\label{app:nonuni}
When there is background cosmic-ray heating (balancing cooling), the linearised equations are
\begin{equation} \label{eq:nonuni_drho}
    \frac{\delta \rho}{\rho} - \frac{3}{2} \eta^{-1} \omega_c \frac{\xi_z}{v_{A,z}} = -i \bm{k \cdot \xi},
\end{equation}
\begin{equation}  \label{eq:nonuni_dxi}
    -\rho \omega^2 \bm{\xi} = -i \bm{k} \Big( \delta p_c + \delta p_g + \frac{\bm{B \cdot \delta B}}{4\pi} \Big) + i \frac{\bm{(B\cdot k) \delta B}}{4\pi} - \omega_{\rm ff} c_s \delta \rho \ \bm{\hat{z}}
\end{equation}
\begin{equation} \label{eq:nonuni_dB}
    \bm{\delta B} = i(\bm{B \cdot k})\bm{\xi} - i \bm{B}(\bm{k \cdot \xi}) ,
\end{equation}
\begin{multline} \label{eq:nonuni_pg}
    \frac{\delta p_g}{p_g} \Big( \frac{\omega}{\gamma - 1} + i\omega_c \LT   \Big) + \frac{\omega \omega_c \xi_z}{(\gamma-1) v_{A,z}} \Big( 1 - \gamma \frac{\omega_{\rm ff}}{\omega_c} \frac{v_{A,z}}{c_s}  \Big)  =   \omega_a \frac{\delta p_c}{p_g} \\
    - \omega_a \omega_c \frac{\xi_z}{v_{A,z}} - i \omega_c \frac{\delta \rho}{\rho}  \Big( \frac{5}{2} - \LT \Big) -  \Big( \frac{\gamma}{\gamma - 1}  \omega + i \omega_c \Big) i \bm{k \cdot \xi} ,
\end{multline}
\begin{multline} \label{eq:nonuni_pc}
    \frac{\delta p_c}{p_g} (\omega - \omega_a + i \omega_d + i \omega_c \eta^{-1}) - \omega \omega_c \frac{\xi_z}{v_{A,z}} = \frac{\delta \rho}{\rho}  \Big( - \frac{2}{3} \eta \omega_a + i \omega_c  \Big) \\ - \frac{4}{3} i \eta \omega \bm{k \cdot \xi}  + i \kappa  \bm{\hat{b} \cdot k} ( \bm{ \delta \hat{b} \cdot \nabla } p_c)  + i \kappa \bm{ k \cdot \delta \hat{b}} ( \bm{ \hat{b} \cdot \nabla } p_c),  
\end{multline}
where $\bm{\delta \hat{b}} = \bm{\delta B} / B - \delta B / B \ \bm{\hat{b}}$. For the calculations in Section \ref{sec:nonuni}, where we ignore gravity, we set $\omega_{\rm ff} = 0$ in the above equations.

\section{Validity of the 1D thermal instability calculation} \label{app:1D_val}
For a high-$\beta$ uniform medium, the dispersion relation of the thermally unstable entropy mode can be derived simply by imposing pressure balance $\delta p_c \simeq - \delta p_g$ and combining the CR and gas energy equations. This leads directly to equation \ref{eq:dispersion_quad_uni}.  The same approach does not work in the presence of background cosmic-ray heating and/or gravity because then the cosmic-ray and gas energy equations have terms proportional to the fluid displacement $\xi$ (see Appendix \ref{app:nonuni}) and so imposing $\delta p_c \simeq - \delta p_g$ is not sufficient to uniquely determine the entropy mode properties.   In this Appendix, we discuss the approximations that successfully reproduce the entropy mode in this limit.   In particular, we explain why the 1D calculation in Section \ref{sec:nonuni_1d}, which assumes $\bm{\xi} \parallel \bm{B}$, is a reasonable approximation for the thermal instability eigenfrequency.  

We note from the start that the usual Boussinesq approximation, $\bm{k \cdot \xi} =0$, often utilized to impose pressure balanced fluctuations, is not appropriate for this problem. Instead, for $\omega_a \gg \omega_c$, pressure balance, $\delta p_c + \delta p_g \approx 0$ ($\omega \ll k c_s$ implies that $\delta p_c + \delta p_g \ll \delta \rho c_s^2$), simply determines the leading-order gas entropy frequency (eq. \ref{eq:uni_dr_stram}), which satisfies $\omega < \omega_a$ for all $\eta$. For $\eta \lesssim 1$, $\omega \ll \omega_a$ and the induction equation implies that $\delta B / B \ll \delta \rho /\rho$. Moreover, one can show that
\begin{equation}
    \frac{\xi_{\perp}}{\xi_{\parallel}} = \frac{k_{\perp} k_{\parallel} \omega^2}{\omega^2 k_{\parallel}^2 - \omega_{a}^2 k^2} \sim \frac{\omega^2}{\omega_a^2},
\end{equation}
where $\xi_\perp$ and $\xi_\parallel$ are the fluid displacements perpendicular and parallel to the magnetic field, respectively.
So, for $\eta \lesssim 1$, where $\omega \ll \omega_a$, we can restrict our analysis to field-aligned perturbations $\bm{\xi} \parallel \bm{B}$, for which $\bm{\delta B} =0$ (just like in the 1D calculation). It turns out that $\bm{\xi} \parallel \bm{B}$ yields the same dispersion relation as the 1D calculation \eqref{eq:disp_1d}, independent of propagation direction (for a fixed $\omega_a$). 

For $\eta \gg 1$ thermal instability becomes isochoric. As a result, the 1D  dispersion relation still gives the correct thermal-instability eigenfrequency, even though $\bm{\xi} \parallel \bm{B}$ is not strictly true ($\xi_\parallel$ still exceeds $\xi_\perp$ by a factor of a few). When $\delta p_c / p_g \gg \delta \rho / \rho$, $\bm{\xi}$ is not important for thermal instability, as it is tied to density perturbations.

The 1D calculation also works well in the limit of strong diffusion, $\omega_d \gg \omega_a$. In this case, the $\omega$ of the gas entropy mode also never exceeds $\omega_a$. This, again, is determined by $\delta p_c + \delta p_g \approx 0$, which to leading order gives the quadratic:
\begin{equation} 
     \eta \Big( \frac{4}{3}\omega - \frac{2}{3}\omega_a   \Big) \Big( \frac{3}{2}\omega +  \omega_a \Big) + \frac{5}{2}  \omega \Big(    \omega - \omega_a + i \omega_d \Big)  = 0.
\end{equation}
Small deviations (primarily in the real part) between the 1D thermal instability eigenfrequency and the exact solution occur only when $\eta \sim \omega_c \omega_d / \omega_a$, where thermal instability is most strongly damped by diffusion. The CR-diffusion induced damping rate is $\mathcal{O}(\omega_a)$, while $\delta p_c / p_g \sim \delta \rho / \rho$, and so the assumption that $\bm{\xi} \parallel \bm{B}$ is only approximately well motivated. We stress, however, that the deviations (which mainly affect the oscillation frequency) occur only in the case where thermal instability is very rapidly damped.

\section{Convective Instability: Growth-rate Derivation from linearised equations} \label{app:conv}
In this section we derive an approximate growth rate for the CR convective instability in the limit $\omega_{\rm ff} \gg \omega_c, \omega_a$. The growth rate is not exact, as we drop any dependence that is $\mathcal{O}(\omega_c) \ll \omega_{\rm ff}, \omega$. We consider the simplest case of a purely vertical magnetic field and horizontal propagation: $\bm{B} =B  \ez$ and $\bm{k} =k  \ex$, for which $\omega_a = 0$.

In this limit, one can derive the approximate dispersion relation
\begin{equation} \label{eq:quadquart}
    \omega^4 + b \omega^2   +c =0,
\end{equation}
where
\begin{equation} \label{eq:b}
    b = - \Big( \omega_s^2 +\frac{4}{3 \gamma} \eta \omega_s^2  + \frac{3}{2}  \eta^{-1} \omega_c \omega_{\rm ff} \frac{c_s}{v_{A,z}} \Big),
\end{equation}
\begin{equation} \label{eq:c}
    c = -   \frac{\omega_s^2 \omega_c \omega_{\rm ff}}{2 \gamma \eta} \frac{c_s}{v_{A,z}} \Big[ 2 \eta  \Big(\gamma \frac{\omega_{\rm ff}  v_{A,z}}{\omega_c c_s} -2 \Big)  - 3 \gamma     \Big].
\end{equation}
This has solutions of the form:
\begin{equation} \label{eq:quart_disp}
\omega^2 = \frac{-b \pm \sqrt{b^2 - 4c}}{2}
\end{equation}
which will have an unstable branch if $c<0$, i.e.
\begin{equation}
2 \eta   \Big(\gamma \frac{\omega_{\rm ff}}{\omega_c}\frac{v_{A,z}}{c_s} -2 \Big) > 3 \gamma ,
\end{equation}
which is the same condition as obtained in the main text using the Schwarzschild criterion (equation \ref{eq:schwarz_cond}). Equations \eqref{eq:b}--\eqref{eq:quart_disp} can be combined to give an expression for the growth rate. In the common limit $b^2 \gg c$ (large $\omega_s$ limit), the growth rate simplifies to
\begin{equation} \label{eq:conv_rate}
\omega \approx i \sqrt{\frac{c}{b}} \approx i \sqrt{\omega_c \omega_{\rm ff} \frac{c_s}{v_{A,z}}} \  \Big[ \  \frac{\frac{1}{\gamma} \Big(\gamma \frac{\omega_{\rm ff}  v_{A,z}}{\omega_c c_s} -2 \Big)  - \frac{3}{2}\eta^{-1} }  {1 + 4 \eta / (3\gamma)} \  \Big]^{1/2}
\end{equation}
Note that as $\eta \rightarrow \infty$, the growth rate goes to 0. The dashed curve in Figure \ref{fig:convection} shows the approximate growth rate from equation \eqref{eq:conv_rate}, which agrees well with the exact calculation.


\bsp	
\label{lastpage}
\end{document}